\documentclass[twocolumn,prd,nofootinbib]{revtex4}
\newcommand\ForRichardOnly[1]{}
\usepackage{verbatim}
\usepackage{color}     
\usepackage{framed}
\definecolor{shadecolor}{gray}{0.95}
\usepackage{amsmath}
\usepackage{graphicx}  
\usepackage{tabularx}
\usepackage{wrapfig}   
\usepackage{hyperref}
\usepackage{booktabs}
\usepackage{subfigure}
\usepackage{xspace}

\usepackage{mathtools}

\newcommand\unit[1]{{\rm #1}}
\newcommand\Y[1]{Y^{(#1)}{}}

\newcommand\mc{{{\cal M}_c}}
\newcommand\msun{$\textrm{M}_{\odot}$}

\newcommand\qmstateproduct[2]{\left\langle#1|#2\right\rangle}

\newcommand\BS{\textsc{bayestar}}
\newcommand\gstlal{\textsc{GSTlal}\xspace}
\newcommand\Like{{\cal L}}
\newcommand\RedLike{{{\cal L}_{\rm red}}}

\newcommand\citeMCMC{\cite{LIGO-CBC-S6-PE,2011PhRvD..83h2002D,2011PhRvD..84f2003C,gr-extensions-tests-Europeans2011,gwastro-mergers-PE-Aylott-LIGOATest,2011ApJ...739...99N,gw-astro-PE-Raymond,PhysRevD.91.042003}}

\newcommand\itrprm{\vec{\lambda}}
\newcommand\etrprm{\vec{\theta}}

\newcommand{\savefootnote}[2]{\footnote{\label{#1}#2}}
\newcommand{\repeatfootnote}[1]{\textsuperscript{\ref{#1}}}

\begin{document}

\title{A novel scheme for rapid parallel parameter estimation of gravitational waves from compact binary coalescences}
\author{C. Pankow}
\email{pankow@gravity.phys.uwm.edu}
\affiliation{Center for Gravitation, Cosmology, and Astrophysics, University of Wisconsin-Milwaukee, Milwaukee, WI 53201, USA }
\author{P. Brady}
\email{patrick@gravity.phys.uwm.edu}
\affiliation{Center for Gravitation, Cosmology, and Astrophysics, University of Wisconsin-Milwaukee, Milwaukee, WI 53201, USA }
\author{E. Ochsner}
\email{evano@gravity.phys.uwm.edu}
\affiliation{Center for Gravitation, Cosmology, and Astrophysics, University of Wisconsin-Milwaukee, Milwaukee, WI 53201, USA }
\author{R. O'Shaughnessy}
\email{oshaughn@mail.rit.edu}
\affiliation{Center for Computational Relativity and Gravitation, Rochester Institute of Technology, Rochester, NY 14623, USA}
\affiliation{Center for Gravitation, Cosmology, and Astrophysics, University of Wisconsin-Milwaukee, Milwaukee, WI 53201, USA }

\begin{abstract}
We introduce a highly-parallelizable architecture for estimating parameters of compact binary coalescence using gravitational-wave data and waveform models. Using a spherical harmonic mode decomposition, the waveform is expressed as a sum over modes that depend on the intrinsic parameters (e.g. masses) with coefficients that depend on the observer dependent extrinsic parameters (e.g. distance, sky position). The data is then prefiltered against those modes, at fixed intrinsic parameters, enabling efficiently evaluation of the likelihood for generic source positions and orientations, independent of waveform length or generation time.  We efficiently parallelize our intrinsic space calculation by integrating over all extrinsic parameters using a Monte Carlo integration strategy. Since the waveform generation and prefiltering happens only once, the cost of integration dominates the procedure. Also, we operate hierarchically, using information from existing gravitational-wave searches to identify the regions of parameter space to emphasize in our sampling. As proof of concept and verification of the result, we have implemented this algorithm using standard time-domain waveforms, processing each event in less than one hour on recent computing hardware. For most events we evaluate the marginalized likelihood (evidence) with statistical errors of $\lesssim 5\%$, and even smaller in many cases. With a bounded runtime independent of the waveform model starting frequency, a nearly-unchanged strategy could estimate NS-NS parameters in the 2018 advanced LIGO era. Our algorithm is usable with any noise curve and existing time-domain model at any mass, including some waveforms which are computationally costly to evolve.
\end{abstract}
\maketitle

\section{Introduction}
\label{sec:introduction}

The upcoming ground based gravitational-wave detector network (notably including advanced LIGO~\cite{0264-9381-27-8-084006} and advanced Virgo~\cite{0264-9381-32-2-024001}) are sensitive to the gravitational-wave signal from coalescing compact binaries, both the relatively well understood signal from the earlier inspiral phase~\cite{2003PhRvD..67j4025B,2004PhRvD..70j4003B,2004PhRvD..70f4028D,BCV:PTF,2005PhRvD..71b4039K,2005PhRvD..72h4027B,2006PhRvD..73l4012K,2007MNRAS.374..721T,2008PhRvD..78j4007H,gr-astro-eccentric-NR-2008,gw-astro-mergers-approximations-SpinningPNHigherHarmonics,gw-astro-PN-Comparison-AlessandraSathya2009} and the less well understood strong-field merger~\cite{2011PhRvD..83l2005A,2009CQGra..26p5008A, 2014PhRvD..89d2002K,2009PhRvD..79l4028B,2010PhRvD..82f4016S,2011CQGra..28m4002M,2013PhRvD..87b4009M,2013CQGra..31b5012H}. Both of these regimes are important in understanding the underlying physical processes and properties of the binary which likely provide the central engine for phenomena such as short gamma-ray bursts~\cite{2009ARAA..47..567G,Gehrels-shortgrb-SwiftReview-Mid2007,2006RPPh...69.2259M}. A multimessenger observation of this category of event would be the first of its kind and could answer open questions about these poorly understood events. In the era of multimessenger astronomy, rapid and robust measurements of candidate compact binary gravitational-wave events will be a critical science product for gravitational-wave observatories, as colleagues with other instruments perform followup and coincident observations~\cite{LIGO-2013-WhitePaper-CoordinatedEMObserving}. The most tantalizing proposed electromagnetic counterparts are expected to be brief, potentially disappearing within days if not shorter~\cite{2012ApJ...746...48M,2014MNRAS.439..757G,2014MNRAS.437L...6K,2014MNRAS.437.1821M,2014ApJ...780...31T,2013ApJ...775...18B,first2years}. Given limited resources, reliable low-latency parameter estimation of gravitational-wave signals will significantly enhance the science output of multimessener astronomy.

With the scheduled resumption of data taking in late 2015~\cite{LIGO-2013-WhitePaper-CoordinatedEMObserving}, the second generation gravitational-wave interferometers in Hanford, Washington, and Livingston, Louisiana, are expected to reach unprecedented sensitivities~\cite{s6sens}. The Virgo detector is also expected to resume data taking within a year of this milestone. In preparation for the next run, the LIGO and Virgo Collaborations have implemented and extensively tested a set of low latency gravitational-wave detection pipelines~\cite{s6lowlatency,s6opt,gstlal,first2years}, capable of compact binary event detection within a few minutes (or less) from the coalescence time. These pipelines trade the ability to accurately determine all but a few of the parameters of the coalescence for speed and breadth of analysis.

Having a more accurate estimation of the binary coalescence's parameters is valuable not only to gravitational-wave science but also  to  electromagnetic observatories to guide their  pointing. Many astrophysical phenomena which could create transient gravitational waves also have electromagnetic signatures which may decay rapidly. Moreover, gravitational-wave interferometer networks often cannot localize gravitational-wave events to better than a few hundred square degrees on the sky~\cite{LIGO-2013-WhitePaper-CoordinatedEMObserving,first2years,gwastro-skyloc-Sidery2013}, implying follow-up observations should occur expeditiously after the initial identification to promptly localize as accurately as possible. While several Bayesian algorithms for gravitational-wave parameter estimation have been employed in the past, the time scale of a full parameter estimation analysis remains at a much higher latency than the initial detection.

With these goals in mind, and moving hierarchically from the broader search result, fast follow-ups have been developed to rapidly localize the event in the sky, assuming the search pipelines have provided accurate timing and masses~\cite{2011CQGra..28j5021F,first2years}. However, the most robust interpretation of gravitational-wave data requires systematically comparing all possible candidate signals to the data, constructing a Bayesian posterior probability distribution for candidate binary parameters~\citeMCMC{}. These schemes provide measurements of all the physical parameters of the coalescence, albeit at a much higher latency than the search itself. Owing to the complexity and multimodality of the model space, these strategies have adapted variants of Markov Chain Monte Carlo (MCMC) or nested sampling~\cite{2011RvMP...83..943V,PhysRevD.91.042003} algorithms to estimate the parameters of the coalescence. In physics, similar path-based methods have been enormously successful at a broad range of physical problems, by exploring all possible paths through a configuration space; see, e.g.~\cite{2001RvMP...73...33F,1987PhLB..195..216D}.

Though successful, these nature of these algorithms is functionally serial, with moderate degrees of parallelization requiring intensive communication to coordinate the current state. Without state-of-the-art techniques like ensemble sampling, MCMC and nested sampling techniques therefore do not scale efficiently to beyond a few tens of cores (e.g., for parallel tempering). Convergence of these algorithms is also limited by \emph{ergodicity}. No theorem guarantees they must explore the entire model space, let alone efficiently; no expression can robustly assess convergence, using available sampled data. In contrast, efficient and highly-parallelizable Monte Carlo integration strategies are frequently applied to problems with dimensions comparable or higher than coalescing compact binaries; see, e.g.,~\cite{lepage1980vegas,1980PhRvD..21.2308C,book-math-Jaeckel-MonteCarlo,mm-QuasiMonteCarlo-Papageorgiou2001}. In this work, we apply such methods to gravitational-wave parameter estimation for the first time.

This work is organized as follows. In Sec. \ref{sec:Executive} we provide an executive summary, outlining the principles that enable our algorithm to provide rapid, accurate results for the test cases explored here: time-domain models for non-spinning, circular binaries. Then, in Sec. \ref{sec:Waveforms} and \ref{sec:Methods}, we outline the pertinent features of our waveform decomposition and describe our algorithm in detail. To demonstrate our algorithm provides high performance in an environment mimicking conditions during an observational run, Sec. \ref{sec:Results} presents results drawn from a large sample of events from the ``2015 double neutron star mock data challenge,'' results from which are described in~\cite{first2years}. In Sec. \ref{sec:Conclude}, we discuss the broader significance of our result in the context of other parameter estimation work inside and outside the community.

\section{Executive Summary}
\label{sec:Executive}

In this paper, we outline an alternative architecture for gravitational-wave parameter estimation, based on three key
ingredients: efficient evaluation of the likelihood as a function of extrinsic parameters, highly-parallelized Monte Carlo integration over a grid of intrinsic parameters, and input derived from gravitational-wave search and sky localization pipelines.  

The computational cost to evaluate the likelihood has historically been a limiting step in gravitational-wave parameter estimation. This cost has two factors in each comparison which is required to be made. The first is the cost of waveform generation, as discussed in~\cite{gwastro-mergers-PE-ReducedOrder-2013,2013PhRvD..87l2002S,2013PhRvD..87d4008C,gwastro-mergers-IMRPhenomP,gwastro-SpinTaylorF2-2013} and references therein. Second is the filtering cost related to the required length of the waveform to be compared --- this scales in proportion to the lowest frequency accessible by a gravitational-wave interferometer. At fixed sampling rates, the lowest frequency content of the waveform dominates a time-domain likelihood computation. 
Recently, several methods have been proposed to perform this comparison more efficiently~\cite{gwastro-mergers-PE-ReducedOrder-2013,2013PhRvD..87l2002S,2013PhRvD..87d4008C,gw-astro-ReducedOrderQuadraturePE-TiglioEtAl2014}, by interpolating some combination of the waveform or likelihood or by adopting a sparse representation to reduce the computational cost of data handling. In this work, we introduce a robust and straightforward scheme to reduce the cost per comparison, without resorting to interpolation within or manipulation of the waveform family. We first perform a ``precomputation'' phase for each intrinsic parameter by decomposing each physically distinct source in the spin-weighted spherical harmonic basis. This allows us to quickly evaluate the likelihood as a function of extrinsic parameters. This method is applicable to any available waveform model, and the computational cost of the precomputation step requires the waveform to be generated and decomposed only once, thereby making this step a small fractional cost of the total computation. 
 To our knowledge, this work is the first time any such method has been implemented for the most accurate but computationally-expensive waveform models like EOB~\cite{gw-astro-EOBspin-Tarrachini2012,gw-astro-EOBNR-Calibrated-2009} while leveraging several hundred computing cores at once.

Taking advantage of the fast likelihood computation, each set of intrinsic parameters has a straightforward and vectorizable Monte Carlo integration performed to explore and marginalize over the extrinsic parameters. Monte Carlo integration can provide an exhaustive search of the parameter space with a fixed cost in convergence scaling with the number of points drawn. As each Monte Carlo draw is a statistically independent sample, this procedure can engage as many computational resources as are necessary to provide fast convergence of the integral to a satisfactory level of uncertainty. 
Efficient Monte Carlo integration schemes adjust their sampling strategy while performing the integration by adapting to sample only the meaningful regions of the space. By using detection search results (e.g. those produced by matched-filtered searches~\cite{ihope}) to sample near maxima of the likelihood, Monte Carlo methods can retain generality but still be extremely efficient. We describe a procedure to explore the extrinsic parameter space and evaluate the marginalized likelihood that is embarrassingly parallel, extensibly incorporates information provided to accelerate convergence, and provides a well-understood sampling error estimate, allowing us to target a given level of precision.

To complete the parameter estimation process, our algorithm evaluates the likelihood and accumulates intrinsic posterior samples using a well-motivated discrete grid whose position and size is derived from the existing information provided by gravitational-wave detection pipelines. To further accelerate convergence, we also use the information provided by the search to prioritize specific combinations of extrinsic parameters for further investigation. Like the process of marginalizing the likelihood at fixed intrinsic parameters, this grid search is embarrassingly parallel.    
Typically, parameter estimation strategies  make almost no use of the information reported by existing search codes, precisely to avoid self-consistency issues that can arise by, for example, using inferences for the data as priors on the reanalysis of that data.  In contrast, by carefully distinguishing between sampling and integration priors, we circumvent these issues.

To summarize, combining these three factors (parallelizable Monte Carlo integration, efficient likelihoods, and search information), we can provide reliable parameter estimates for merging double neutron star binaries within one hour on existing hardware for instruments available in the next five years.
This promising and simple first implementation  has performance comparable to or better than well-developed state-of-the-art Markov Chain Monte Carlo codes like \textsc{lalinference} applied to initial detector data with comparable nonprecessing sources. For advanced instruments, with longer signals and more costly waveforms, our method should perform significantly better than the current \textsc{lalinference} implementation with state-of-the-art nonprecessing binaries. Our method therefore provides a valuable alternative to other promising methods like reduced-order modeling, ensemble sampling, and other tactics to improve scaling and decrease latency \cite{2013APS..APRG10003F,2014PhRvD..90b4014F}. Owing to its superior scaling and transparency, our method provides a well understood and easily implemented alternative to \textsc{lalinference} for suitable problems, particularly valuable for low-latency and approximate parameter estimation and for investigations into waveform systematics.

\section{Binary Waveforms}
\label{sec:Waveforms}
\subsection{Intrinsic and extrinsic parameters}

On physical grounds, we group waveform parameters $\vec{\mu}=(\vec{\lambda},\vec{\theta})$ into two classes: the intrinsic parameters ($\vec{\lambda}$) and extrinsic parameters ($\vec{\theta}$). The intrinsic parameters are fundamental to the description of the binary: if we change any intrinsic parameters we must recompute the orbital dynamics of the binary (typically through the relatively expensive process of numerically integrating ordinary differential equations). Extrinsic parameters simply describe how the binary is oriented in space and time relative to the detector; changing extrinsic parameters involves a relatively inexpensive rotation, translation or rescaling transformation. As we will show in the next subsection, for the non-spinning case considered here the intrinsic parameters (transformations of the binary component masses $m_1$ and $m_2$) are

\begin{equation} \label{eq:intrinsic}
\itrprm=\{\mc,\eta\}\ ,
\end{equation}

\noindent where $\mc = (m_1m_2)^{3/5}/(m_1+m_2)^{1/5}$ is the chirp mass and $\eta = m_1m_2/(m_1+m_2)^2$ is the symmetric mass ratio. The extrinsic parameters are

\begin{equation} \label{eq:extrinsic}
\etrprm=\{t_{\rm geo},\alpha,\delta,\iota,D,\psi,\phi_c\}\ ,
\end{equation}

\noindent where $t_{\rm geo}$ is the time at which the coalescence point of the waveform arrives at the Earth geocenter, $\alpha$ and $\delta$ are the right ascension and declination, $\iota$ is the inclination angle of the binary's angular momentum vector and the line of sight to Earth, $D$ is the luminosity distance to the binary\footnote{For the sources considered in this paper, the redshift correction is assumed to be negligible.}, $\psi$ is the polarization angle, and $\phi_c$ is the orbital phase of the binary at coalescence.

\subsection{Waveform decomposition}

The gravitational wave strain measured by the $k^{\rm th}$ interferometric detector in a network is given by

\begin{equation} \label{eq:measured_strain}
h_k(t) = F_{+,k}(\delta, \alpha, \psi) h_{+,k}(t) 
+ F_{\times,k}(\delta, \alpha, \psi) h_{\times,k}(t)\ ,
\end{equation}

\noindent where $F_{+,k}$, $F_{\times,k}$ are the antenna patterns of the detector\footnote{Due to the rotation of the Earth, the antenna patterns change as a function of time. While this has been mostly neglected due to the signal duration versus the Earth's rotational velocity, an accounting of this effect will become necessary as the instruments become sensitive to longer binary coalescence waveforms.} and $h_{+,k}$, $h_{\times,k}$ are the two components of the gravitational wave strain, evaluated at the $k^{\rm th}$ detector. The antenna patterns depend only on the extrinsic sky location and polarization angle, while the polarizations depend on both intrinsic and extrinsic parameters. Meanwhile, at leading order for inspiral-only waveforms the polarizations are described $\Phi(t)$ the orbital phase, and the post-Newtonian $v(t)$ ``velocity'' parameters, these depending only on combinations of the masses of the binary. Thus, it is convenient to introduce $h(t|\mc,\eta,\iota,\phi_c)$ given by:

\begin{align}
h_{+,k}(t)-i h_{\times, k}(t) &= h(t-t_k|\mc,\eta,\iota,\phi_c)\ .
\end{align}

\noindent In this expression, $t_k$  denotes the time of arrival of the coalescence  at the $k^{th}$ detector,

\begin{equation} \label{eq:t_k}
t_k = t_{\rm geo} - \frac{\vec{x}_k \cdot \hat{N}(\alpha,\delta)}{c}\ ,
\end{equation}

\noindent where $\vec{x}_k$ is a vector pointing from the geocenter to the $k^{th}$ detector and $\hat{N}(\alpha,\delta)$ is the direction of gravitational-wave propagation. So, each member of the network will have an offset relative to the geocenter time depending only on sky location.

At this stage, we make no assumptions about the functional form of $h(t|\mc,\eta,\iota,\phi_c)$. We can use a $-2$ spin-weighted spherical harmonic mode decomposition (denoted $h_{lm}$) to further separate intrinsic and extrinsic parameters appearing in the polarizations as

\begin{widetext}
\begin{equation} \label{eq:h:Expansion}
 h_{+,k}(t) - i\,h_{\times,k}(t)=\frac{D_{\rm ref}}{D} \sum_{lm} \hat{h}_{lm}(\mc,\eta,t_k;t) \Y{-2}_{lm}(\iota,-\phi_c),
\end{equation}
\end{widetext}

\noindent evaluated at some fixed distance $D_{\rm ref}$ (in this work we choose $D_{\rm ref} = 100$ Mpc).

If we define a complex-valued antenna pattern for each detector as

\begin{equation} \label{eq:complexF}
F_k = F_{+,k} + i \, F_{\times,k} \ ,
\end{equation}

\noindent then we can re-express the measured strain in the $k^{th}$ detector as

\begin{widetext}
\begin{equation} \label{eq:decomposed_strain}
h_k(\itrprm,\etrprm;t) = {\rm Re}\ \frac{D_{\rm ref}}{D}\, F_k(\alpha,\delta,\psi) \, \sum_{lm} 
\hat{h}_{lm}(\mc,\eta,t_k;t)\; \Y{-2}_{lm}(\iota,-\phi_c)\ .
\end{equation}
\end{widetext}

\noindent Aside from $t_k$, we have now completely separated the intrinsic parameters (which enter only the $\hat{h}_{lm}$) from the extrinsic parameters (which enter only the $F_k$ and $\Y{-2}_{lm}$).

Given a time-domain representation of the gravitational wave strain, we can define a frequency-domain version of this strain via a Fourier transform

\begin{equation} \label{eq:Fourier}
\tilde{h}(f) = \int_{-\infty}^\infty h(t) e^{- 2 \pi i f t} dt\ .
\end{equation}

\noindent Time translation of frequency-domain waveforms is trivial. If $\tilde{h}(t_k;f)$ is the Fourier-domain representation of a strain for some arrival time $t_k$, then the same strain arriving at another time $t_k'$ can be simply related by

\begin{equation} \label{eq:time_shift}
\tilde{h}(t_k';f) = \tilde{h}(t_k;f) \, e^{- 2 \pi i f (t_k' - t_k)}\ .
\end{equation}

\noindent Thus, if we work with frequency-domain waveforms, the arrival time $t_k$ can be factored out as $\exp( - 2 \pi i f t_k)$ and we can complete the separation of intrinsic and extrinsic parameters.

For the explicit decomposition above, we have focused on non-spinning waveforms and found that they can be separated into two intrinsic parameters and seven extrinsic parameters. In the general case, which could include precession, tidal effects and any other physics, this intrinsic and extrinsic separation is still possible. In fact, there will always be seven extrinsic parameters which enter only through inexpensive geometric factors, while any additional parameters will always be encoded in the expensive $h_{lm}$ modes.

To see that this is true, first note that Eq.~(\ref{eq:time_shift}) holds for an arbitrary strain and so can always be used to factor out the dependence on time of arrival. Similarly, a gravitational wave far from its source will always fall off as $1/D$. Furthermore, the antenna patterns $F_+$ and $F_\times$ depend on the detector geometry, not the source, and so are unchanged. This gives a total of five extrinsic parameters that in general can be factored out exactly as in our non-spinning waveform. Lastly, we have two angles which enter into the $\Y{-2}_{lm}$. The physical interpretation of these angles depends on the choice of the frame relative to which the harmonic mode decomposition is performed. One should choose a frame which is convenient for expressing and computing the $h_{lm}$. For the non-spinning case, this means aligning the frame with the orbital angular momentum, $\hat{L}$, in which case one can show that the zenith angle is $\iota$ and the azimuth is $- \phi_c$. In the precessing case, one would most likely use a frame aligned with the total angular momentum, $\hat{J}$, e.g. using the parameterization described in~\cite{Farr:2014qka}. In the notation of that paper, the extrinsic spherical harmonic dependence is $\Y{-2}_{lm}(\theta_{JN},-\phi_{JL})$, where $\theta_{JN}$ is the inclination of the \emph{total} angular momentum to line of sight, and $\phi_{JL}$ marks the azimuthal position of $\hat{L}$ on its precession cone about $\hat{J}$.

\section{Methods}
\label{sec:Methods}

A set of data $d_k(t)$ collected from a gravitational-wave detector is typically decomposed into the inherent noise of the detector and a putative gravitational-wave signal:

\begin{equation}
d_k(t) = h_k(t) + n_k(t)\ .
\end{equation}

\noindent In the absence of a signal and under the assumption that each detector produces stationary Gaussian noise, the noise $\tilde{n}_k(f)$ in the $k^{\rm th}$ detector is characterized by its power spectrum $S_k(f)$:

\begin{align}
\left<\tilde{n}_k(f)^* \tilde{n}_k(f')\right> &= \frac{1}{2} S_k(|f|) \delta(f-f')\ .
\end{align}

\noindent In the following discussion, we define a weighted inner-product of two complex Fourier-domain functions $\tilde{a}(f), \tilde{b}(f)$ with a weighting function $1/S(f)$:

\begin{align} \label{eq:InnerProduct}
\qmstateproduct{a}{b} \equiv 2 \int_{-\infty}^{\infty} df \frac{\tilde{a}^*(f)\tilde{b}(f)}{S(f)}\ .
\end{align}

\noindent The mode decomposed waveforms $\hat{h}_{lm}$, can be shifted in time in reference to a detector relative to the geocentric time via Eq.~(\ref{eq:time_shift}).  For this reason, the overlap between a single detector data time series $d(t)$ and a time-shifted complex function $h(t-t_k)$ is 

\begin{equation} \label{eq:IFFT}
\langle h(t_k) | d \rangle = 2\int_{-\infty}^{\infty} \frac{\tilde{d}(f) \tilde{h}^*(f)}{S(f)} e^{2\pi i f t_k} df \ .
\end{equation}

\noindent Note that this is simply the Fourier transform of the integrand in Eq.~(\ref{eq:InnerProduct}), which means we can compute the overlap for all possible time shifts with a single Fourier transform.

\subsection{Bayesian evidence and posteriors}

If the detector noise is Gaussian, it follows that the probability of some set of noise realizations in each of our detectors is 

\begin{equation}
p(\{d\}|{\cal H}_0) \propto \prod_k \exp \left(- \frac{ \langle d | d \rangle_k}{2}\right)
\end{equation}

\noindent in the absence of a gravitational-wave signal. Here, ${\cal H}_0$ indicates the hypothesis that the data is only Gaussian noise. 

With a gravitational-wave signal present, the data from each detector is noise plus the response of an interferometer to the gravitational-wave $h(\vec{\mu})$. In this case, the probability of some measured $d$ given the presence of the signal with parameter values $\vec{\mu}$ is 

\begin{equation}
p(\{d\}|\vec{\mu},{\cal H}_1) \propto \prod_k \exp \left(- \frac{ \langle d - h(\vec{\mu}) | d - h(\vec{\mu}) \rangle_k }{2} \right)
\end{equation}

\noindent where ${\cal H}_1$ indicates the hypothesis that the data consists of Gaussian noise plus a gravitational-wave signal. 

By Bayes' theorem, the posterior probability of the parameters $\vec{\mu}$ under the signal present hypothesis ${\cal H}_1$ is

\begin{equation}
\label{eqn:bayes}
p(\vec{\mu}, {\cal H}_1 | \{d\}) = \frac{p(\{d\} | \vec{\mu}, {\cal H}_1) p(\vec{\mu})}{ p(\{d\} | {\cal H}_1) }
\end{equation}

\noindent where

\begin{equation}
p(\{d\} | {\cal H}_1) = \int p(\{d\} | \vec{\mu}, {\cal H}_1) p(\vec{\mu}) d\vec{\mu} \ .
\end{equation}

\noindent It is convenient, at this stage, to introduce the likelihood ratio

\begin{equation} 
\label{eq:likelihood}
\Like(\vec{\mu}|\{d\}) = \prod_k \frac{\exp \left\{ - \langle d - h(\vec{\mu}) | d - h(\vec{\mu}) \rangle_k / 2 \right\} } {\exp \left\{ - \langle d | d \rangle_k / 2 \right\} }
\end{equation}

\noindent and to rewrite Eq.~(\ref{eqn:bayes}) in the form

\begin{equation}
p(\vec{\mu}, {\cal H}_1 | \{d\}) = \frac{\mathcal{L}(\vec{\mu}|\{d\}) p(\vec{\mu})}{Z} \ ,
\end{equation}

\noindent where $Z$ is defined to be

\begin{equation}
\label{eq:def:Z:Modified}
Z = \int \mathcal{L}(\vec{\mu}|\{d\}) p(\vec{\mu}) d\vec{\mu} \equiv \frac{p(\{d\} |{\cal H}_1)}{p(\{d\} | {\cal H}_0)} \ .
\end{equation}

\noindent We can also compute the posterior probability density (which we will abbreviate as simply the posterior) for one or multiple parameters (marginalized over all other parameters). Let $x$ be one or more parameters in $\vec{\mu}$ and $y$ be all other parameters, such that $\vec{\mu} = x \cup y$. Then, the posterior for $x$ is

\begin{equation} \label{eq:def:Posterior}
p({\cal H}_1|x) \equiv \frac{p(x|{\cal H}_1)}{Z(\{d\}|{\cal H}_1)} \ \int dy \;  p(y|{\cal H}_1) \Like(x,y)\ .
\end{equation}

\subsection{Efficiently evaluating the likelihood}

We now show how the waveform decomposition derived above can be exploited to speed up evaluations of the likelihood ratio. In Eq.~(\ref{eq:decomposed_strain}) note that we have taken the observed strain in a detector $H_k(t)$ and rewritten it as a linear combination of harmonic mode time series $\hat{h}_{lm}(t)$. The harmonic mode time series depend on the intrinsic parameters, but the extrinsic parameters (apart from $t_k$) are entirely encoded in the coefficients of the linear combination. Because computing the likelihood is just an inner product, which is a linear operator, we can pull these coefficients outside the inner product integral. Thus, we need only compute inner products involving the $\hat{h}_{lm}$ and data $\{d\}$. If we store these, we can compute the likelihood for any extrinsic parameters by simply recomputing the coefficients and reconstructing the linear combination.

To that end, we define the following quantities:

\begin{subequations}
\label{eq:QUV}
\begin{align}
Q_{k,lm}(\itrprm,t_k) &\equiv \qmstateproduct{h_{lm}(\itrprm,t_k)}{d}_k \nonumber\\
&= 2 \int_{-\infty}^{\infty} \frac{df}{S_k(|f|)} e^{2\pi i f t_k} \tilde{h}_{lm}^*(\itrprm;f) \tilde{d}(f)\ , \\
{ U_{k,lm,l'm'}(\itrprm)} &= \qmstateproduct{h_{lm}}{h_{l'm'}}_k\ , \\
V_{k,lm,l'm'}(\itrprm) &= \qmstateproduct{h_{lm}^*}{h_{l'm'}}_k  \ .
\end{align}
\end{subequations}

\noindent Note that each $Q_{k,lm}(\itrprm,t_k)$ is computed for all $t_k$ with a single inverse Fourier transform, as in Eq.~(\ref{eq:IFFT}). A signal will produce a peak in the filtered outputs localized to a short millisecond time window around the coalescence time, so the $Q_{k,lm}(\itrprm,t_k)$ will be sharply peaked as functions of $t_k$ and we need only retain the values for a narrow range of $t_k$. To allow for detector arrival times that differ from the geocenter time, the range of $t_k$ for which we must store the $Q_{k,lm}$ is set by the light travel time across Earth ($2R_{\oplus}/c\simeq 42 \unit{ms}$). Conservatively, work we store the $Q_{k,lm}$ for a much longer $300$ ms range of $t_k$. The $U_{k,lm,l'm'}(\itrprm)$ and $V_{k,lm,l'm'}(\itrprm)$ are independent of $t_k$ and are computed once by a straightforward use of Eq.~(\ref{eq:InnerProduct}).

By plugging Eq.~(\ref{eq:decomposed_strain}) into Eq.~(\ref{eq:likelihood}), taking a log and collecting terms, we obtain

\begin{widetext}
\begin{eqnarray}
\ln \Like(\itrprm; \etrprm) 
&=& (D_{\rm ref}/D) \text{Re} \sum_k \sum_{lm}(F_k \Y{-2}_{lm})^* Q_{k,lm}(\itrprm,t_k)\nonumber \\
 &&  -\frac{(D_{\rm ref}/D)^2}{4}\sum_k \sum_{lml'm'}
\left[
{
|F_k|^2 [\Y{-2}_{lm}]^*\Y{-2}_{l'm'} U_{k,lm,l'm'}(\itrprm)
}
 {
+  \text{Re} \left( F_k^2 \Y{-2}_{lm} \Y{-2}_{l'm'} V_{k,lm,l'm'}(\itrprm) \right)
}
\right].\nonumber\\
\label{eq:def:lnL:Decomposed}
\end{eqnarray}
\end{widetext}

\noindent Importantly, the intrinsic parameters $\itrprm$ enter only through the $Q_{k,lm}$, $U_{k,lm,l'm'}$ and $V_{k,lm,l'm'}$. These are the dominant cost, as they require computing the orbital dynamics, the $h_{lm}$, inner product integrals and inverse Fourier transforms. By contrast, the extrinsic parameters enter the $F_k$ and $\Y{-2}_{lm}$, which are much cheaper to compute.

Therefore, if we fix a point in the intrinsic parameter space we can compute the
$Q_{k,lm}(\itrprm,t_k)$, $U_{k,lm,l'm'}(\itrprm)$ and $V_{k,lm,l'm'}(\itrprm)$ only once, vary the extrinsic parameters and compute the likelihood for only the cost of the $F_k$ and $\Y{-2}_{lm}$. This allows us to efficiently integrate over the extrinsic parameters and obtain a marginalized posterior for the intrinsic parameters $p(\itrprm)$ as in Eq.~(\ref{eq:def:Posterior}). If we do this for a collection of points in the intrinsic parameter space, we can integrate over $\itrprm$ as well and obtain $Z$ as in Eq.~(\ref{eq:def:Z:Modified}). Note that the computation for each point in the intrinsic space is completely independent of the others. This makes the algorithm embarrassingly parallel and given enough CPU cores the entire analysis can be run in the time it takes to integrate over the extrinsic parameters (modulo some brief startup and post-processing steps).

The remainder of this section provides more details on the various steps of our algorithm.

\subsection{Placement over intrinsic parameters}
\label{sec:itr_placement}

\begin{figure}
\includegraphics[width=\columnwidth]{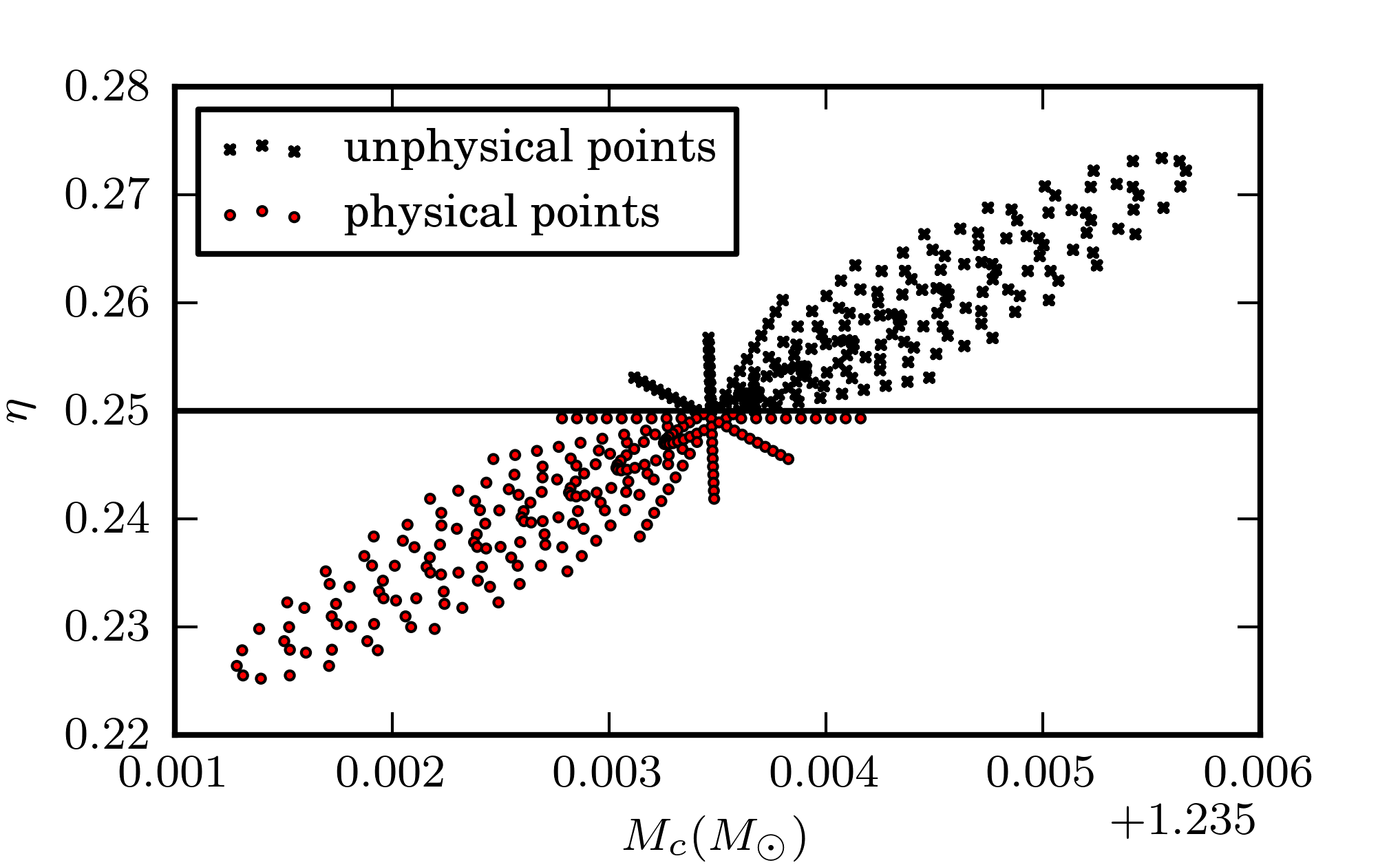}
\caption{\label{fig:linear_ellipse} We use an effective Fisher matrix to compute
an approximate ellipsoidal region of overlap $\geq 90\%$ with the masses reported by a detection pipeline. We then fill this ellipsoid with discrete points and cut any with unphysical values of symmetric mass ratio $\eta$. In this case $m_1= 1.5 {\rm M}_\odot$ and $m_2 = 1.35 {\rm M}_\odot$, so a little under half of the points placed would have been unphysical. At each physical grid point, we marginalize the likelihood over all extrinsic parameters as described in Sec.~\ref{subsec:extrinsic}.}
\end{figure}

\noindent In the current work, we restrict ourselves to consider non-spinning binaries in which we also neglect tidal effects. Therefore, we need only two mass parameters to describe the intrinsic parameter space, for which we use the symmetric mass ratio $\eta = m_1 m_2 / M^2$ and the chirp mass ${\cal M}_c = M \eta^{3/5}$ (where $M = m_1+m_2$ is the total mass). Since detection searches will report masses for the candidate event, we use these to guide which region of the intrinsic parameter space to explore.

Let $\left( {\cal M}^*,\eta^* \right)$ be the masses reported by a detection pipeline. We then perform an effective Fisher matrix calculation as described in~\cite{gwastro-mergers-HeeSuk-FisherMatrixWithAmplitudeCorrections,gwastro-mergers-HeeSuk-CompareToPE-Aligned} centered about this point. This involves evaluating the overlap 

\begin{equation}
\label{eqn:overlap}
\frac{|\langle h(\itrprm^*)|h(\itrprm)\rangle|}{\sqrt{\langle h(\itrprm^*)|h(\itrprm^*)\rangle\langle h(\itrprm)|h(\itrprm)\rangle}}
\end{equation}

\noindent between our waveform with masses $\left( {\cal M}^*,\eta^* \right)$ and approximately tens of nearby waveforms with different intrinsic parameters (while extrinsic parameters are held constant). The measured overlap values are then fit with a multi-dimensional quadratic. The coefficients of this quadratic fit are called the effective Fisher matrix. Like the standard Fisher matrix, the effective Fisher matrix serves as a quick, crude estimate of expected parameter estimation performance and can be used to predict surfaces of constant overlap, which will in general be ellipsoids. In this work, we use the effective Fisher matrix to approximate the region of intrinsic parameter space which will have overlap $\geq 90\%$ with the masses $\left( {\cal M}^*,\eta^* \right)$.

Once we have defined this $90\%$ overlap ellipsoid, we must fill it with a set of discrete points at which we will compute the likelihood. To do this, we first specify the total number of intrinsic parameter points we wish to place, here  200 points. Then, these points are arranged within a unit sphere. In this work, we placed points along 20 radial lines, with 10 points per spoke. Along each spoke, the points are placed uniformly in radial distance. Now, the eigenvalues and eigenvectors of the effective Fisher matrix tell us the lengths and orientations of the axes of the $90\%$ overlap ellipsoid. We use these to deform and rotate our set of points in the sphere to a set of points in the $90\%$ overlap ellipsoid. Once we have filled our ellipsoid, we remove any points that have unphysical $\eta$ ($>1/4$). We ensure that we always place spokes in our ellipsoid along the direction of constant $\eta$, so that we always have many points along this boundary of the parameter space. One reason for the choice of spoked placement is so that we can always ensure near-equal mass binaries will have many points near the $\eta = 0.25$ boundary of the parameter space, the region where most astrophysical binary neutron stars are expected. Fig.~\ref{fig:linear_ellipse} illustrates this placement of intrinsic parameter points.

\subsection{Integrating over extrinsic parameters}
\label{subsec:extrinsic}

Because precomputed quantities allow us to efficiently evaluate the likelihood $\Like(\itrprm,\etrprm)$ as a function of $\etrprm$, we first integrate the likelihood $\Like(\itrprm,\etrprm)$ over all extrinsic parameters $\etrprm$ to get

\begin{align}
\RedLike(\itrprm) &= \int \Like(\itrprm,\etrprm) p(\etrprm) d\etrprm\ ,
\end{align}

\noindent where $p(\etrprm)$ is our prior over the extrinsic  parameters.  We assume the sources analyzed are randomly-oriented and randomly distributed in the universe out to a fiducial radius\footnote{Beyond which we do not expect our instruments to be appreciably sensitive in the next two years.}. With the potential exception of the sky position, our priors are independent, and thus separable. We then evaluate the reduced likelihood $\RedLike$ by integrating over the geocentric time of arrival $t$, distance $D$, sky position $\Omega_{\rm sky}$ represented as right ascension $\alpha$ and declination $\delta$, angular momentum orientation as measured by inclination $\iota$, polarization angle $\psi$, and coalescence phase $\phi_{\rm c}$, using the separable prior to get

\begin{align}
\RedLike(\itrprm) &=  \nonumber \\
 & \int \frac{dt}{T_{\rm window}} \frac{D^2 dD d\Omega_{\rm sky} }{V_{\rm max}} \frac{d (\cos \iota) d\phi_c}{4\pi} \frac{d\psi}{\pi} p(\etrprm) \Like(\itrprm,\etrprm)\ .
\end{align}

\noindent In this expression and our calculations, we adopt a maximum distance $D_{\rm max}$ (here $300~\unit{Mpc}$, beyond which we do not expect appreciable sensitivity from a 2015 era interferometer) and a time window $ T_{\rm window}$  (here, $300~\unit{ms}$) surrounding the event.  

With the exception of time (described below), we evaluate these integrals and reconstruct the posterior distribution using Monte Carlo integration \cite{book-mm-NumericalRecipies,peter1978new}. If $p_s(\etrprm)$ is a distribution which is never zero, then 

\begin{align}
\RedLike(\itrprm) &= \int \frac{\Like(\itrprm,\etrprm) p(\etrprm)}{p_s(\etrprm)} [p_s(\etrprm) d\etrprm]\ .
\end{align}

\noindent If we draw $N$ random samples $\etrprm_q$ from $p_s$, we can estimate the value of the integral $\hat{{\cal L}}_{\rm red}$ and its error $\sigma_{\RedLike}$ as follows. Let

\begin{align}
w_q &\equiv \frac{\Like(\itrprm,\etrprm_q) p(\etrprm_q)}{p_s(\etrprm_q)},
\end{align}

\noindent for a given draw $q$ of $\etrprm$, then

\begin{align}
\label{eqn:mc_int_mean}
\hat{{\cal L}}_{\rm red}(\itrprm) &\equiv \frac{1}{N} \sum_q w_q = \left<w\right>, \\
\sigma_{\RedLike}^2 &= \left<w^2\right> - \left<w\right>^2\ .
\label{eqn:int_uncert}
\end{align}

The weighted samples also provide an estimate of the marginalized one-dimensional cumulative distributions $\hat{P}(<x)$ at fixed $\itrprm$:

\begin{align}
\hat{P}(<x) &\equiv \frac{1}{\sum\limits_q w_q} \sum\limits_q w_q \Theta(x-x_q),
\end{align}

\noindent where $x$ is one of the extrinsic variables in $\etrprm$, and $\Theta$ is the Heaviside step function. This formula follows by performing Monte Carlo integration on the parameter volume $<x$, keeping track of overall normalization. In the limit of many samples, this discontinuous estimate should converge to a smooth, marginalized posterior distribution.  
For any set of samples and any $x$, the error in $\hat{P}$ follows from the (correlated) statistics of the Monte Carlo integrals in its numerator and denominator. In the typical case that all samples $x_q$ are distinct, the unique sample with the largest weight corresponds to the largest discontinuity in $\hat{P}$. The magnitude of this discontinuity, or equivalently its inverse $n_{\rm eff}$, provides a practical measure of how reliable we expect this one-dimensional posterior to be:

\begin{align}
n_{\rm eff} &\equiv \frac{\sum\limits_q w_q}{\max \{w_{\rm q}\}}\ .
\label{eqn:neff}
\end{align}

\noindent Equivalently, the ``effective number of samples'' $n_{\rm eff}$ measures how many independent samples produce similar weights near the largest observed weight. 

Following the discussion of priors earlier in this section, unless otherwise indicated, we draw samples using a separable sampling distribution $p_s(\etrprm) = \prod_{\alpha}p_{s,\alpha}(\theta_\alpha)$. Each factor $p_{s,\alpha}$ is equal to the corresponding prior in that dimension. 

Furthermore, being a pure Monte Carlo integral, all the draws are independent. We can simply average the results of multiple instances to achieve a smaller $\sigma_{\RedLike}$, even if these evaluations adopted to a different sampling distribution. Currently, we perform $n_{\rm trials}$ ($=10$) evaluations of the integral at each fixed intrinsic point, terminating when either $N$ iterations crosses a threshold ($=10^6$) or to some fixed $n_{\rm eff}$ threshold ($=1000$), whichever comes first. This approach is thus highly parallelizable: simply instantiate instances (with different seeds) of the desired integral across however many computing resources are required for the target execution time and precision, effectively dividing the time required to converge by the number of processes available. Combined with gridding the intrinsic parameters, this allows for a degree of parallelization which has not been achieved by other sampling methods in this parameter space. There is a practical limit to this strategy since there is a fixed start up time, but the degree of parallelization realized is at least an order of magnitude or more over current schemes.

\subsubsection{Adaptive Monte Carlo integration}

To better concentrate our samples in regions of high significance, we implemented an importance sampling algorithm using a simple adaptive Monte Carlo procedure. Every $n_{\rm adapt}$ samples, we updated the sampling distribution based on measured weights $w_k^{\beta}$. We choose to temper the distribution by raising the likelihood to an exponent which is chosen heuristically based on the network SNR and the particulars of the adaptive histogram. This choice attempts to mitigate the large dynamic range of $\mathcal{L}$ in Eq. \ref{eqn:mc_int_mean} to a scale comparable to the number of samples used in each histogram.  
At the end of every adaptation period, we reconstruct the one-dimensional sampling distributions in each adapting dimension, using the last $n_{\rm adapt}$ samples. In each dimension ($\theta$), we subdivide the full range into $n_{\rm bins}$ equal-length bins then evaluate a tempered, weighted histogram. Since the dimensionality of the space is high, but sparse, there is a danger that early adaptations will have one-dimensional sampling distributions which suffer from fluctuations in other parameters after marginalization. To avoid this effect and better ensure the full prior space is covered adequately, we then average the histogram $W$ with a uniform distribution with weight $s$:

\begin{align}
\hat{W}_\alpha &= s W_\alpha + (1-s)/n_{\rm bins}\ .
\end{align}

\noindent Finally, we transformed from this discrete, bin-by-bin representation to a continuous integral by constructing a one-dimensional sampling distribution $p_{s,\alpha}(\theta_{\alpha})$ by interpolating between bin centers, then constructing the one-dimensional inverse $P_{s,\alpha}(<\theta_{\alpha})^{-1}$ by integrating $P_{s,\alpha}'(\theta_{\alpha})=p_{s,\alpha}(\theta_{\alpha})$. The latter process ensures that the sampling distribution and inverse cumulative distribution used to generate random samples are normalized and self-consistent.   

As configured for this paper, we used $n_{\rm bins}=100, n_{\rm adapt}=1000$, and $s=0.1$. The adapted sampling distributions are frozen in after $10^5$ samples are drawn.

The adaptive sampler was not used for all parameters. The parameters that are useful to adapt often have a significant dynamic range (i.e., distance) or could be represented with more detailed prior information (i.e. sky location). Conversely, parameters whose marginalized posterior is expected to resemble the prior, like orbital phase and polarization, were not adapted as our one-dimensional method would provide no benefit.  

\subsubsection{Using search results to target specific areas of the sky}

Gravitational-wave search pipelines have usually already identified candidate event times in two or more interferometers' data, typically to much less than the light crossing time of the earth. 

However, the initial search pipeline is not designed to provide estimates of the sky location, so the \BS{} pipeline \cite{singer_thesis} rapidly processes the results of a gravitational-wave search to identify candidate sky locations consistent with a gravitational-wave event, using only the information provided by the top-level search. This code produces a posterior distribution $p_{BS}(\alpha,\delta)$ for the gravitational-wave signal sky location in a discretized, interlocking, equal-area grid of pixels corresponding to sky regions, with probabilities for each pixel (a HEALPix\footnote{\url{http://healpix.sourceforge.net/}} skymap). The refinement of the grid is variable and scales with the resolution required to resolve the features of the posterior in $\alpha$ and $\delta$.  

While we could allow the adaptive sampling to reduce the sky area required to be searched, we can also use the two-dimensional $(\alpha, \delta)$ posterior from \BS{} as a sampling distribution to immediately target our Monte Carlo at a region of high support. This is more effective than the adaptive sampler because the one-dimensional sampling distributions, even if well converged to the true marginalized one-dimensional posteriors, over cover the two-dimensional space, thus limiting their effectiveness in sampling only where support truly exists.

For simplicity, when using a \BS{} skymap, we adopt a purely discrete sky: the samples are selected from the set of sky region centers as calculated by the healpix library. In practice, the pixels are at a sufficient resolution such that detrimental effects from the discretization are rarely noticeable.

\subsubsection{Time marginalization}
\label{sec:time_marg}

Having already computed the $Q_{k,lm}(\vec{\lambda}, t_k)$, we can evaluate the likelihood versus time $\Like(t)$ cheaply, by array addition operations. Hence, rather than performing Monte Carlo integration, we can likewise efficiently \emph{integrate} over time.  

Specifically, for every sky location drawn by the adaptive Monte Carlo routine, we compute the corresponding time shift between geocenter and each detector with Eq.~(\ref{eq:t_k}). To evaluate the value of $\ln {\cal L}$ for some geocenter time $t_{\rm geo}$, we simply look up the corresponding value of the $Q_{k,lm}(\itrprm,t_k)$ from our precomputed values and plug them into Eq.~(\ref{eq:def:lnL:Decomposed}).  
The resulting time series $\Like(t)$ is then numerically integrated using Simpson's rule in a window of 300 ms, centered on $t_{\rm geo}$.

\subsection{Intrinsic priors and posterior construction}
\label{eqn:mcetaprior}

At this point, we have evaluated $\RedLike(\itrprm)$ over a structured grid of intrinsic parameters (here indexed by $r$) $\lambda_r$. By construction, the values $\RedLike(\itrprm)$ have small statistical errors (e.g., typically less than a few percent). We can then interpolate $\RedLike(\itrprm)$ throughout the sampled grid. Combined with the prior $p(\itrprm)$ over intrinsic parameters, we evaluate the overall evidence $Z$ and posterior distribution $p(\itrprm)$ over intrinsic parameters via

\begin{align}
Z &= \int p(\itrprm) \RedLike(\itrprm) d\itrprm\ ,\\
p(\itrprm) &= \frac{1}{Z} p(\itrprm) \RedLike(\itrprm)\ .
\end{align}

To construct posteriors for the intrinsic parameters $\mc,\eta$, we adopt a uniform prior in $m_1,m_2$ with $m_1,m_2\in[1,2]M_\odot$: inside the specified region, the (uniform) prior density is $p(m_1,m_2)dm_1dm_2=  dm_1 dm_2/( M_\odot)^2$.  Changing coordinates using the Jacobian $d(m_1,m_2)/d(\mc,\eta)= \delta \mc/M^2 = \delta \eta^{6/5}\mc^{-1}$ where $\delta = (m_1-m_2)/M = \sqrt{1-4\eta}$, we find the prior density in $\mc,\eta$ coordinates is

\begin{align}
p(\mc,\eta) d\mc d\eta &=  \frac{1}{ M_\odot^2} \frac{\mc d\mc d\eta}{\eta^{6/5}\sqrt{1-4\eta}}\ .
\end{align}

\noindent In other words, due to a coordinate singularity, the prior in $\eta$ diverges near the equal-mass line. This coordinate singularity has a disproportionate impact on comparable-mass binary posteriors.

However, unlike MCMC and nested-sampling codes, we can reweight our results, allowing the user to reprocess the posterior using any user-specified intrinsic-parameter prior. As a result, our approach also enables several other calculations of practical interest: reanalysis given alternative astrophysical priors, simultaneous analysis of multiple events, adapting the ``prior'' intrinsic (mass, tide~\cite{2014arXiv1402.5156W,2013PhRvD..88d4042R}, cosmological parameter~\cite{2010ApJ...725..496N,2012PhRvD..85b3535T,2012PhRvL.108i1101M}) distribution to reproduce multiple observations, and simultaneous independent constraints from multimessenger observations. 

To a good approximation, the intrinsic and extrinsic parameter distributions often separate after marginalizing in time. In other words, after marginalizing in time, the extrinsic parameter distributions are nearly independent of $\itrprm$. In that case, each individual extrinsic parameter distribution provides a reliable estimate of the posterior. As a crude approximation to the one-dimensional cumulative posterior distribution $P(\theta)$, we simply average all weighted samples $(x_{r,q},w_{r,q})$ from all mass points:

\begin{align}
\hat{P}(<x) &= \frac{\sum\limits_{r,q} w_{r,q}\Theta(x-x_{r,q})}{\sum\limits_{r,q} w_{r,q}}\ ,
\end{align}

\noindent where $\Theta$ is the Heaviside step function.

\section{Results: Runtime details and large scale tests}
\label{sec:Results}

We present a proof-of-principle study of our parameter estimation pipeline, employed in a environment which should resemble the steps taken after the identification of search triggers. To this purpose, a mock data challenge was constructed to exercise both the low latency gravitational wave searches as well as the parameter estimation follow-up processes expected to be applied to gravitational-wave candidates identified by the search \cite{first2years}. This challenge proceeded in several stages, each designed to emulate the anticipated workflow in a low latency binary neutron-star (BNS) detection environment. 

From the reported  search pipeline results, we use the $\mc$ and $\eta$ coordinates from which the intrinsic search space is extrapolated, and the window for time marginalization (see Sec. \ref{sec:time_marg}) is formed around the reported coalescence time $t_r$.  The search pipeline also provides reference power spectral densities $S(f)$ utilized in evaluating the likelihood.

Unless otherwise stated, the likelihoods were evaluated using nonspinning TaylorT4 templates at 3.5 post-Newtonian (PN) order, including only the $(2,\pm 2)$ and $(2,0)$ modes. As we will discuss at length below, this signal model does \emph{not} include all degrees of freedom permitted to the binary in the data. When evaluating the likelihood, waveforms started at $f_{\rm low}=40\unit{Hz}$. The inner product integration Eq. (\ref{eq:InnerProduct}) uses an inverse power spectrum filter targeting the frequency range $[f_{\rm min},f_{\rm max}]=[f_{\rm min}, 2000\unit{Hz}]$, constructed from the measured power spectrum.

Only distance used an adaptive sampling function, starting initially with a constant function. Our distance prior was uniform in volume out to $D=300\unit{Mpc}$. The declination and right ascension were sampled from the posterior provided by \BS. As used here, the skymap had $12\times 64^2$ pixels, roughly one per square degree.

\subsection{Ensemble of events}

We first present the agglomerated results from 450 events drawn randomly from a set of recovered events in \cite{first2years,first2yearsweb}. These events were also processed using \BS and a subset were processed by the \textsc{lalinference} Monte Carlo sampler codes. All events have false alarm rate --- as determined by the \gstlal{} search pipeline --- smaller than one per century. This threshold is motivated by the selection criteria outlined in \cite{LIGO-2013-WhitePaper-CoordinatedEMObserving}.   As in \cite{first2years}, this data set adopts a noise PSD set at   the median sensitivity from fiducial predictions for 2015 LIGO observing \cite{LIGO-Inspiral-Rates,LIGO-2013-WhitePaper-CoordinatedEMObserving}. It is expected that the Virgo interferometer will not be functional during this time period, leaving a two detector LIGO site network. 

A population of gravitational-wave signals from BNSs was added into the data, as described in \cite{first2years} and reviewed here. Events in this set (the 2015 MDC data) were distributed isotropically on the sky, and uniformly in volume out to 219 \unit{Mpc}. The BNS injections had uniform random component masses in $1.2 M_\odot-1.6 M_\odot$ and randomly oriented spins with the dimensionless magnitude not exceeding 0.05. These (precessing) binary signals  were generated via precessing SpinTaylorT4 templates at 3.5 PN order \cite{BCV:PTF}, including all known post-Newtonian modes \cite{gw-astro-mergers-approximations-SpinningPNHigherHarmonics}.  

First, the \gstlal{} BNS search was performed over the MDC set, identifying events for further follow up. For each event identified by the gravitational-wave search pipeline, we distributed the intrinsic points according to the procedure described in Sec. \ref{sec:itr_placement} and evaluated the Monte Carlo integral $10$ times for each mass point, combining each of the runs together after. We outline the form of the priors and sampling functions for each parameter in Table \ref{tbl:priors}.

\begin{table*}
\begin{tabular}{r|c|c}
\hline \\
parameter & physical prior & initial sampling distribution \\
\hline \\
distance ($D$) & uniform in volume & uniform in distance \\
inclination ($\iota$) & uniform in $\cos\iota$, $\iota$ between $(-\pi, \pi)$ & same \\
polarization ($\psi$) & uniform in $(0, 2\pi)$ & same \\
orbital phase ($\phi$) & uniform in $(0, 2\pi)$ & same \\
declination ($\delta$) & uniform in $\cos\delta$, $\delta$ between $(-\pi, \pi)$ & drawn from \BS{} skymap \\
right ascension ($\alpha$) & uniform in $(0, 2\pi)$ & drawn from \BS{} skymap \\
\hline
\end{tabular}
\caption{\label{tbl:priors}Prior distributions used for the extrinsic parameters in the 2015 MDC study.}
\end{table*}

In addition to the information provided by the search pipeline, we also make use of the posterior probability of $\alpha$ and $\delta$, provided by \BS. It is expected that such information will be available within a few minutes of trigger identification. The gains in time to convergence in the Monte Carlo integral are expected to outweigh the loss of time waiting for \BS{} to finish processing if the \BS{} posterior is used as a sampling function for the sky location.

\begin{figure}
\includegraphics[width=\columnwidth]{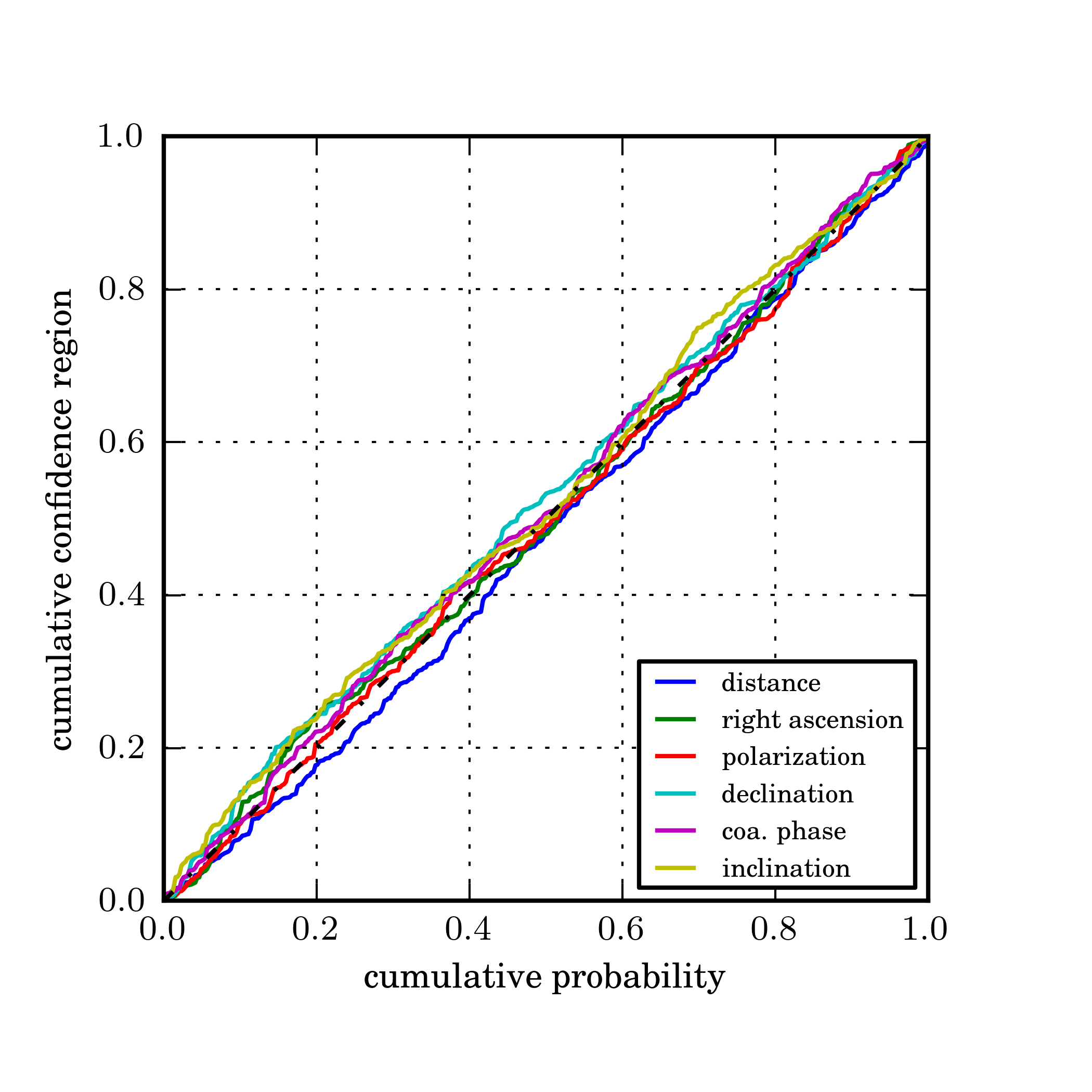}
\caption{\label{fig:pp:2015Ensemble} For 450 randomly-selected NS-NS binaries, a plot of the cumulative distribution of $P_\theta(\theta_k)$ for each extrinsic variable $\theta= D, \alpha, \delta, \iota, \psi, \phi$. The curves represent the normalized fraction of events with estimated cumulative probability less than the injected parameter value.}
\end{figure}

In order to validate the results of the pipeline, we present a comparison of the ensemble distribution of recovered parameters versus their intrinsic distribution as determined by the prior --- a so called PP plot --- in Fig. \ref{fig:pp:2015Ensemble}. If our estimates for the one-dimensional cumulative distributions $P(<x)$ are unbiased and if $x_*$ is a random variable consistent with the prior, then $P(x_*)$ should be a uniformly-distributed random variable. To test this hypothesis, we use the one-dimensional posteriors provided by the MDC.

For each parameter $x$, each colored curve in Fig. \ref{fig:pp:2015Ensemble} is  the fraction of events with estimated cumulative probability $P(<x_*)$ at the injected parameter value $x_*$. Specifically, if $P(x_{*q})$ are the sorted cumulative probabilities for the $q=1\ldots n$ events with $P(x_{*1})<P(x_{*2})$, then the points on the plot are $\{P(x_{*q}),q/n\}$.  

There are a few possible sources of bias in the recovery of the posterior distribution. While the spins are nominally mild, it is possible that the use of non-spinning templates in the signal analysis for both the \gstlal and our parameter estimation codes could cause biases. Additionally, the event selection process outlined in \cite{first2years} introduces a small selection (Malmquist) bias which slightly disfavors edge-on binaries relative to our uniform prior.  Our parameter estimation strategy does not account for selection biases; for a sufficiently large ensemble of events, small deviations between the statistical properties of our posteriors and the ensemble are expected.

\subsection{Marginalization details}

Two events were selected for demonstration. One event represents a marginal detection by the \gstlal pipeline. It is suggestive of what might be seen in the 2015-2016 era, and is not a confident detection by itself. The second event is a ``gold-plated'' event, one which is an assured single-event detection by the \gstlal pipeline. They are qualitatively comparable in most regards, however, the stronger event (\#21091), being more strongly peaked in likelihood did not accumulate as many effective samples, and hence the error on the reduced likelihood ($\RedLike(\lambda)$) tends to be larger by an order of magnitude, yet still small in the relative sense. We provide the simulation and \gstlal event IDs so that the interested reader can cross-correlate results here with those obtained by other samplers in \cite{first2years}. We will refer to them by their \gstlal IDs \#21091 and \#14631. Their properties are enumerated along with the values obtained by the \gstlal pipeline and the recovered parameters in Table \ref{tbl:event_params}.

\begin{table*}

\begin{minipage}[t]{.5\linewidth}

\begin{tabular}{l|ccc}
\hline
\multicolumn{4}{c}{\gstlal Search ID \#21091, Injection ID \#26465} \\
Parameter & Injected & \shortstack{Recovered\\(\gstlal search)} & \shortstack{Recovered\\(parameter estimation)\\median / mode} \\
\hline  
\hline
$m_1$ (\msun) & 1.60 & 1.68 & 1.55 / 1.55 \\
$m_2$ (\msun) & 1.40 & 1.33 & 1.54 / 1.54 \\
$D$ (Mpc) & 68 & - & 59 / 67 \\
$\iota$ & 3.0 & - & 1.3 / 0.47 \\
$\delta$ & -0.12 & - & -0.06 / -1.02 \\
$\alpha$ & 4.00 & - & 3.9 / 0.82 \\
$\psi$ & 3.12 & - & 3.08 / 1.01 \\
$\phi$ & 4.03 & - & 3.08 / 0.94 \\
\hline 
\multicolumn{4}{c}{Other parameters} \\
Parameter & Injected & \shortstack{Recovered\\(\gstlal search)} & \shortstack{Recovered\\(parameter estimation)} \\
\hline
\hline
$|\chi_1|$ & 0.026 & - & - \\
$|\chi_2|$ & 0.008 & - & - \\
$\rho$ & 16.28 & 17.24 \savefootnote{note1}{with corresponding false alarm rate $3.8\times10^{-14}$} & 16.85\savefootnote{maxlnL}{as measured by $2\sqrt{\ln(\max\{L_i\})}$} \\
\hline
\end{tabular}

\end{minipage}%
\begin{minipage}[t]{.5\textwidth}

\begin{tabular}{l|ccc}
\hline
\multicolumn{4}{c}{\gstlal Search ID \#14631, Injection ID \#30639} \\
Parameter & Injected & \shortstack{Recovered\\(\gstlal search)} & \shortstack{Recovered\\(parameter estimation)\\median / mode} \\
\hline  
\hline
$m_1$ (\msun) & 1.32 & 1.49 & 1.49 / 1.49 \\
$m_2$ (\msun) & 1.28 & 1.14 & 1.13 / 1.13 \\
$D$ (Mpc) & 102 & - & 85 / 101 \\
$\iota$ & 3.09 & - & 1.53 / 2.76 \\
$\delta$ & 0.28 & - & 0.28 / 0.91 \\
$\alpha$ & 1.89 & - & 2.39 / 1.83 \\
$\psi$ & 2.32 & - & 3.14 / 3.33 \\
$\phi$ & 5.73 & - & 3.08 / 2.26 \\
\hline 
\multicolumn{4}{c}{Other parameters} \\
Parameter & Injected & \shortstack{Recovered\\(\gstlal search)} & \shortstack{Recovered\\(parameter estimation)} \\
\hline
\hline
$|\chi_1|$ & 0.032 & - & - \\
$|\chi_2|$ & 0.036 & - & - \\
$\rho$ & 11.06 & 12.04\repeatfootnote{note1} & 11.07\repeatfootnote{maxlnL} \\
\hline
\end{tabular}

\end{minipage}

\label{tbl:event_params}
\caption{Injected and recovered intrinsic and extrinsic parameters for an injected signal. The \gstlal search only reports time of arrival, signal-to-noise ratio, and mass information. The parameters from our algorithm are quoted at a weighted median and mode of the marginalized posteriors. Some other parameters not considered in this study (e.g. the dimensionless component spin $\chi_{1,2}$) are listed separately.}
\end{table*}

Each event was gridded in the elliptical region of 10\% mismmatch in the same manner described in Sec. \ref{sec:itr_placement}. After unphysical points were thrown out, there were 128 points for event \#14631 and 124 points for event \#21091. Each point had ten independent instances of the integrator applied, both to accumulate samples in parallel fashion as well as to cross-validate the results. The approximate wall clock time for any given instance was about 45 minutes.

\begin{figure*}
    \centering
    \begin{subfigure}[]{
        \label{fig:21091integral}
        \includegraphics[width=0.95\textwidth]{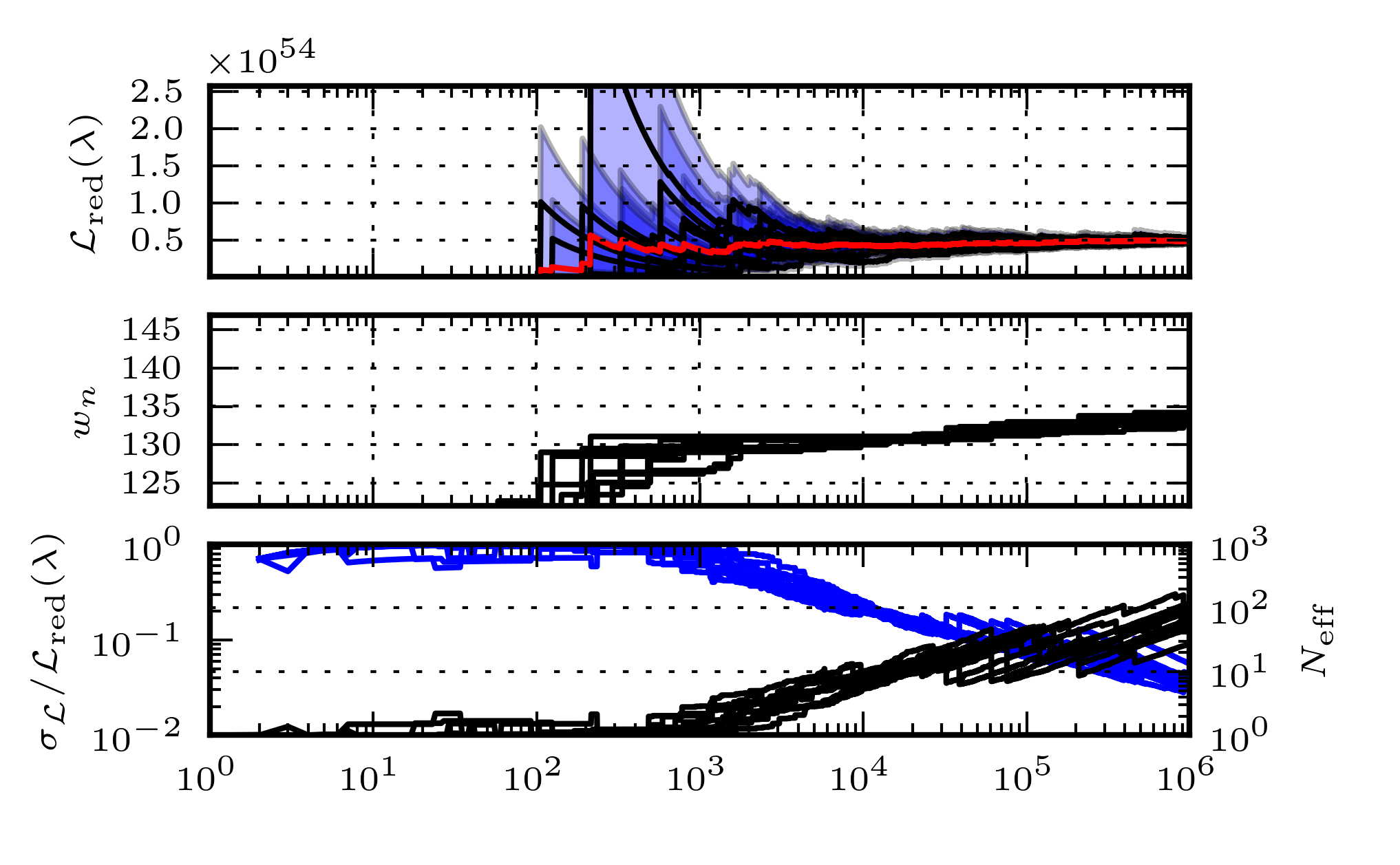}
    }
    \end{subfigure}
    \begin{subfigure}[]{
        \label{fig:14631integral}
        \includegraphics[width=0.95\textwidth]{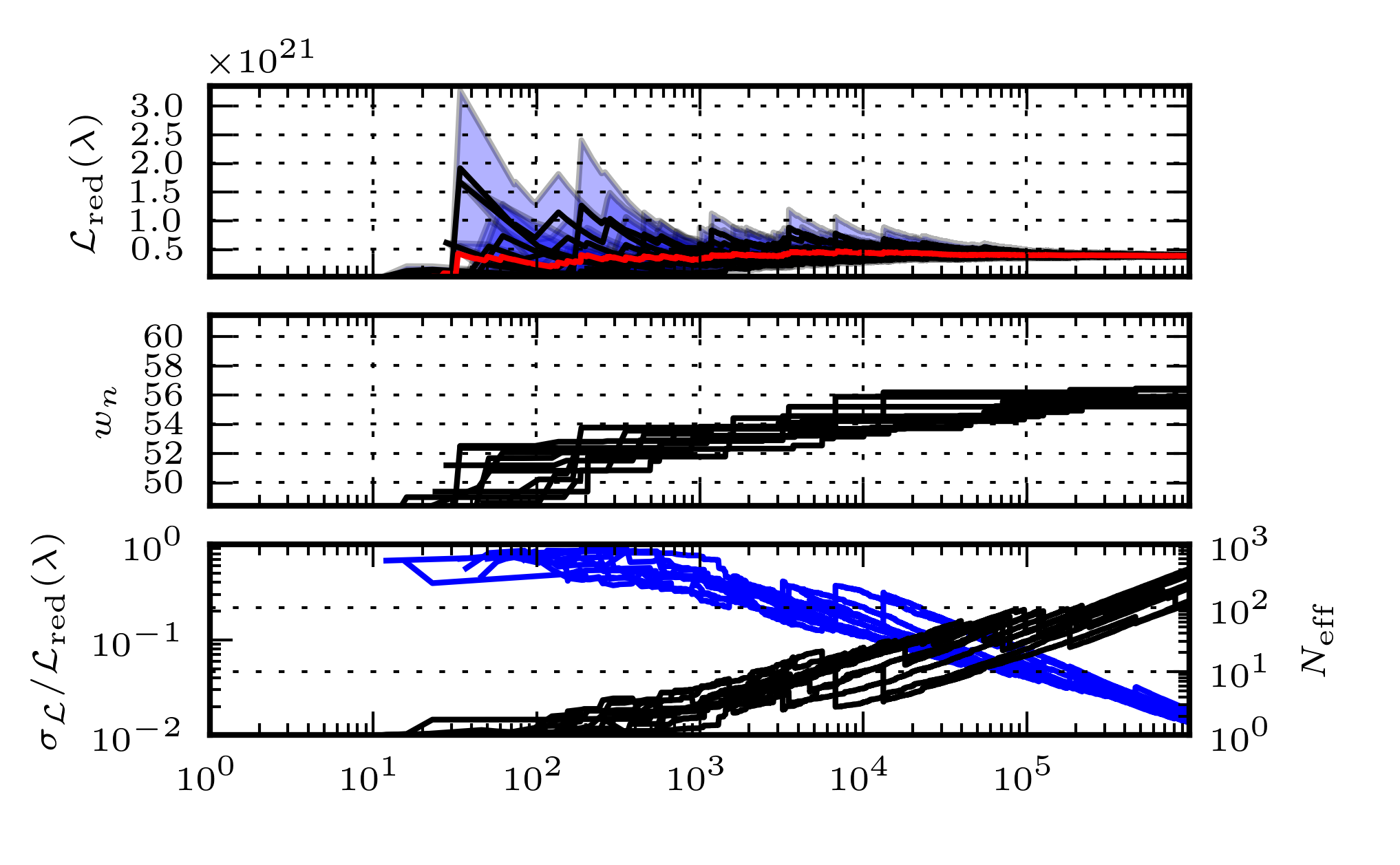}
    }
    \end{subfigure}
\caption{
Figs. (a) and (b) are correspond to events \#21091 and \#14631, respectively. The top panel in each figure shows the estimated value (black lines) of the integral, for each of the ten integrators, through a given sample count. 1-$\sigma$ error regions are shaded in blue around the curve. Additionally, the value of the integral if all ten integrators are combined is shown in red. The middle panel shows the maximum value found by the integrator instance throughout the iterations. The last panel plots the number of effective samples (Eq. (\ref{eqn:neff})) in black, with corresponding ordinate axis the right. In the same panel, the relative error (Eq. (\ref{eqn:int_uncert})) is plotted in blue, with corresponding ordinate axis on the left.
}
\end{figure*}

The evolution of the Monte Carlo integral evaluation is shown in Figs. \ref{fig:21091integral} (\#21091) and \ref{fig:14631integral} (\#14631). This is for a single mass point corresponding to the center of the ellipsoid constructed for sampling the $M_c, \eta$ plane, and so also the mass reported by the \gstlal search. As explained in section \ref{sec:itr_placement} each mass point is independently evaluated 10 times. Each integral evaluation is shown in the top panel as a function of the number of samples drawn to that point. The blue regions surrounding each line are the error estimate at that point in the sampling. It is readily apparent that the integrators have sampled near the maximum point within a few thousand samples (middle panel, Figs. \ref{fig:21091integral} and \ref{fig:14631integral}), and at this point the integrators all begin to converge towards the same evidence value. Within $10^5$ samples, most of the integrators have highly overlapping error regions, and only moderately small changes in the maximum likelihood point searched. The relative error for each integrator is plotted in blue in the bottom panel, with the typical $1/\sqrt{N}$ behavior exhibited. The ``jumps'' in the curves correspond to a new maximum being found in the integrand. The final relative error for any of the integrators is typically of order 5\% or less for all cases (with only one 10\% relative error for \#21091), and typically less than 1\% for \#14631. The red line in the top panel of the figures shows the evidence value if all ten integrators were joined as one integrator.

\begin{figure*}
    \centering
    \begin{subfigure}[]{
        \label{fig:21091intresult}
        \includegraphics[width=0.95\textwidth]{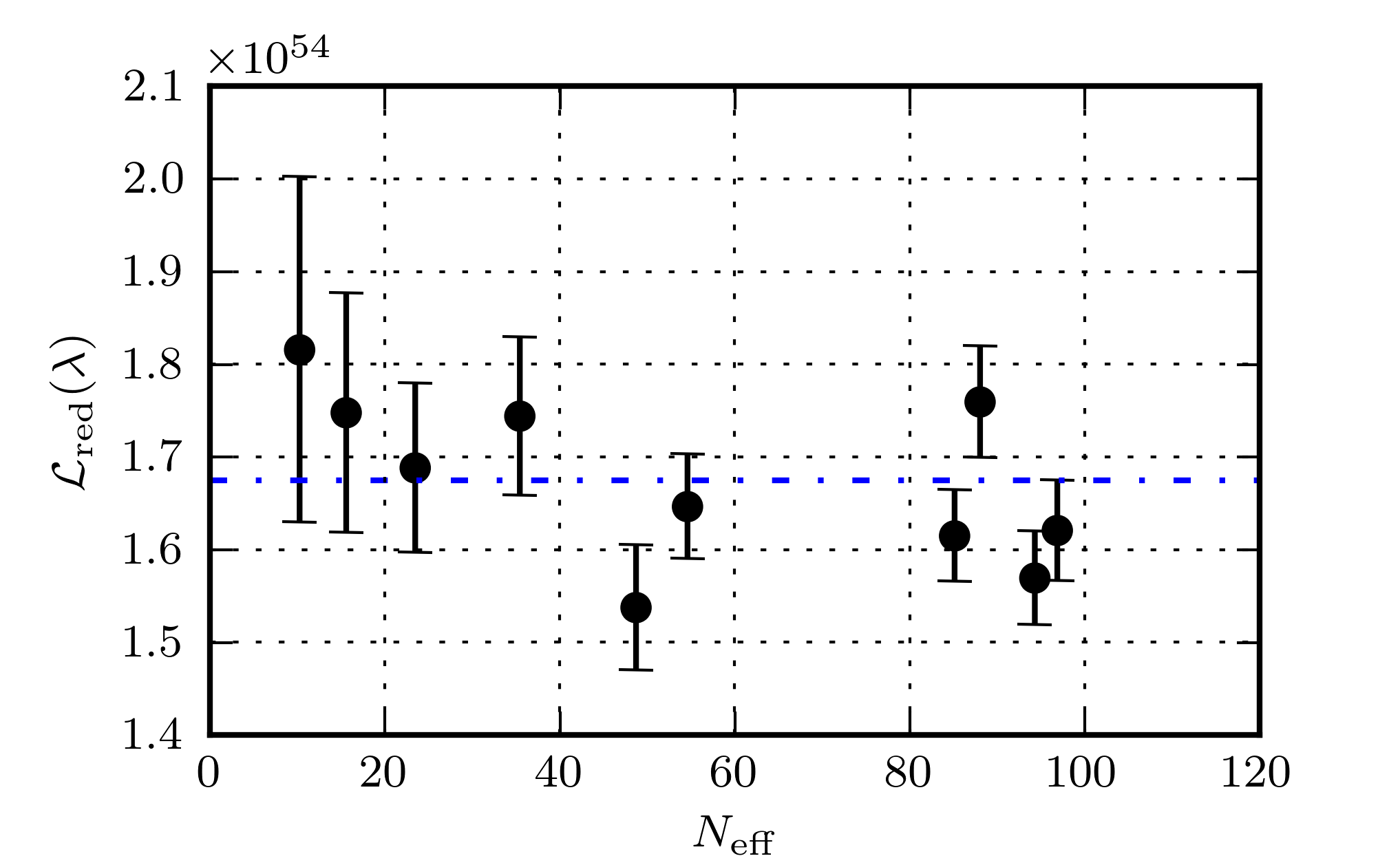}
    }
    \end{subfigure}
    \begin{subfigure}[]{
        \label{fig:14631intresult}
        \includegraphics[width=0.95\textwidth]{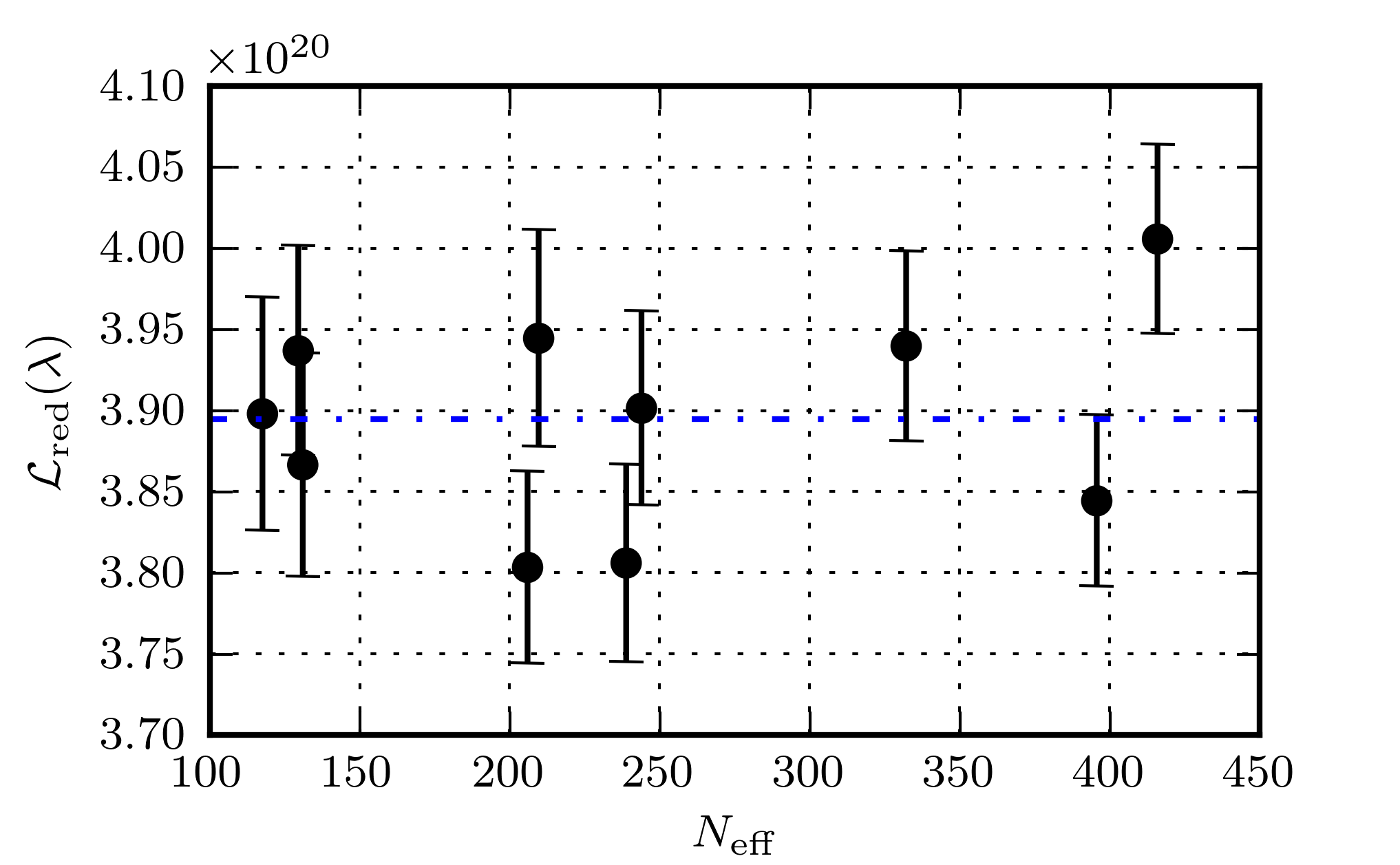}
    }
    \end{subfigure}
\caption{Figs. (a) and (b) correspond to events \#21091 and \#14631, respectively. Each of the 10 integrator results is plotted against the reduced likelihood and number of effective samples collected ($N_{\text{eff}}$). The error bars represent the uncertainty in the integral value from statistical error. The horizontal blue dashed line represents the value of the evidence averaged over all 10 integrator estimates.}
\end{figure*}

\begin{figure*}
    \centering
    \begin{subfigure}[]{
        \label{fig:21091logevid}
        \includegraphics[width=0.95\textwidth]{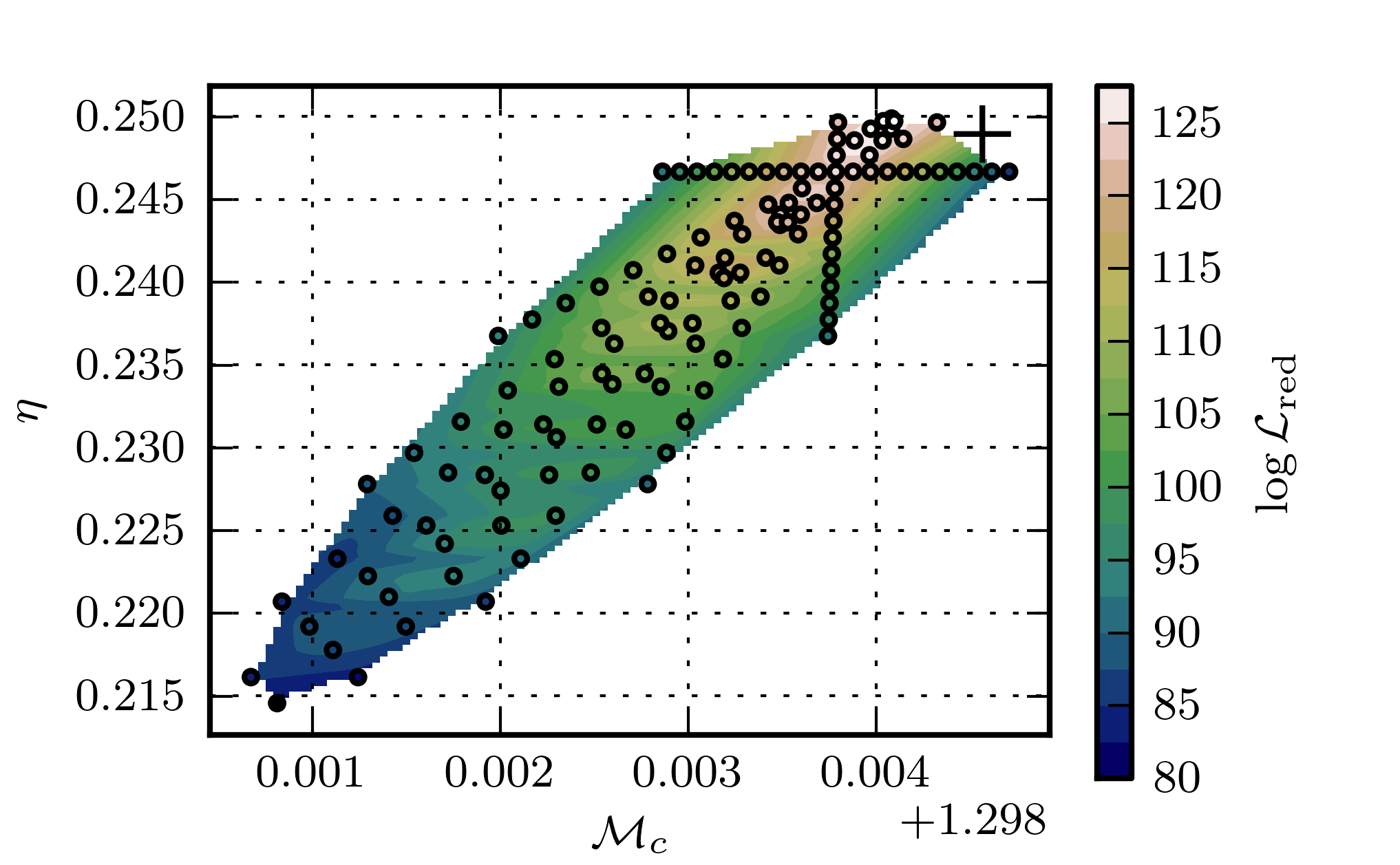}
    }
    \end{subfigure}
    \begin{subfigure}[]{
        \label{fig:14631logevid}
        \includegraphics[width=0.95\textwidth]{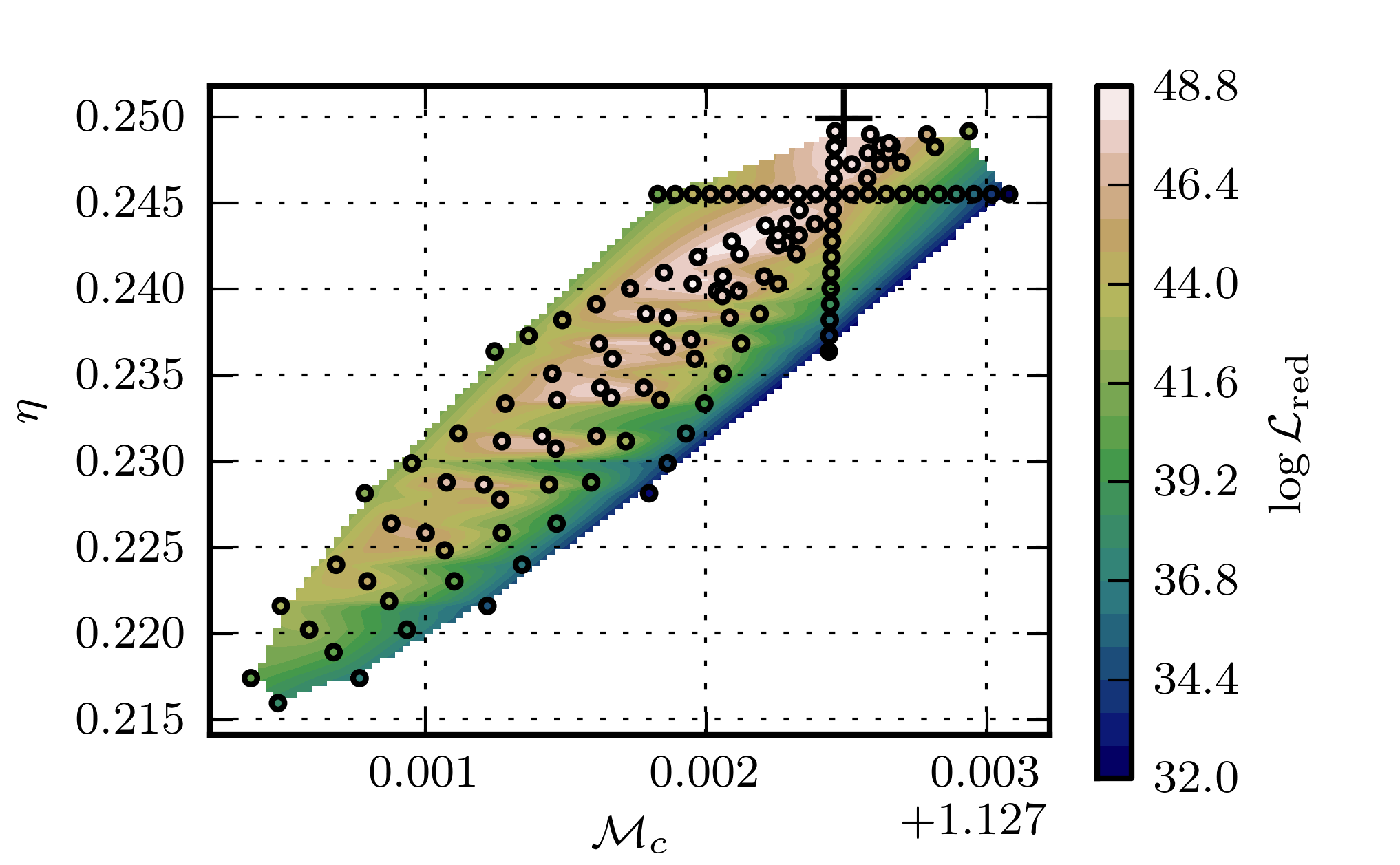}
    }
    \end{subfigure}
\caption{Figs. (a) and (b) correspond to events \#21091 and \#14631, respectively. Level contours of the interpolated evidence surfaces in $\mc$ and $\eta$ space. The points are the values of $\mc, \eta$, which were used to do the integral over the extrinsic parameters, and all 10 integrator instances have been averaged to get the final value. The injected value is marked with a large cross.}
\end{figure*}

Figs. \ref{fig:21091intresult} (\#21091) and \ref{fig:14631intresult} (\#14631) display the results of each of the ten integrator instances scattered in the $N_{\text{eff}}$, evidence plane. The dashed line indicates the value of the evidence as calculated across all samples from all ten points. As can be noted, the error on the evidence values are small and most of the points are clustered near the total evidence dashed line, further indicating that all ten of the integrators have reached an internally consistent value, with very little spread around the average. These evidences are then averaged together to obtain the reduced likelihood for that mass point. A contour plot of the evidence for this event is shown in Figs. \ref{fig:21091logevid} (\#21091) and \ref{fig:14631logevid} (\#14631), with the actual value of $\mc$ and $\eta$ of the injected signal marked by the crosshairs and the value obtained by the \gstlal search pipeline is at the center of the ellipse. The evidence follows roughly the expected quadratic-shape of a Fisher-matrix manifold for these two parameters, but the maximum evidence point is slightly offset from the true value. We expect that this difference arises from the mild spin of the system biasing the result, as the waveform family used to generate templates and calculate $\RedLike$ does not include the effects of spin.

\begin{figure*}
    \centering
    \begin{subfigure}[]{
        \label{fig:21091triplot}
        \includegraphics[width=0.80\textwidth]{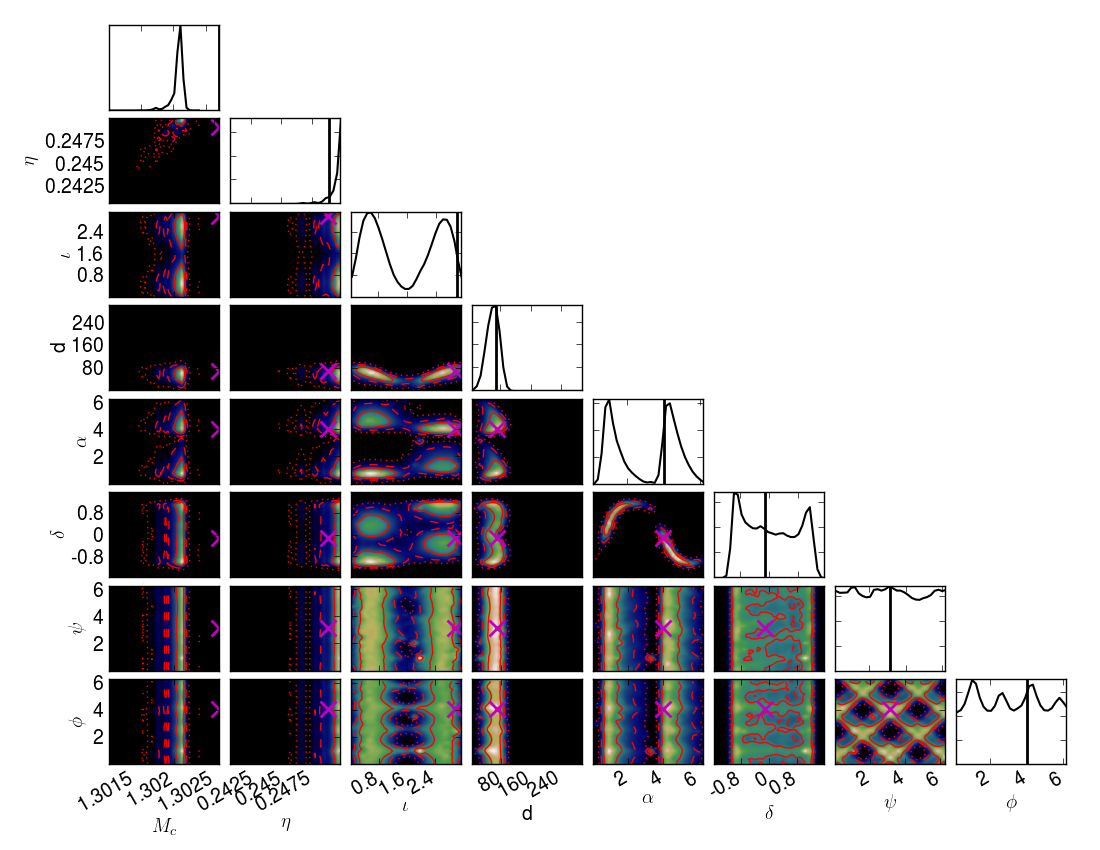}
    }
    \end{subfigure}
    \begin{subfigure}[]{
        \label{fig:14631triplot}
        \includegraphics[width=0.80\textwidth]{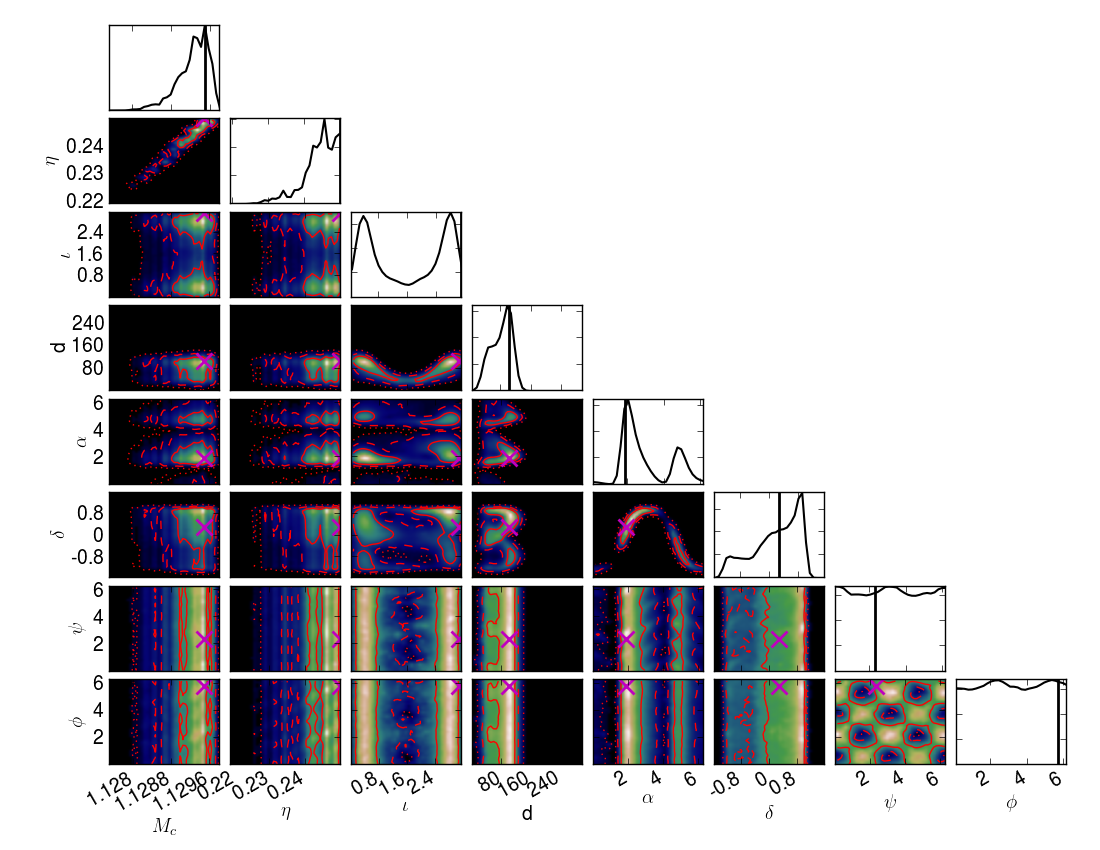}
    }
    \end{subfigure}
\caption{Posteriors for event \#21091, left and event \#14631, right. All two-dimensional posteriors are represented as off-diagonal elements, and the diagonal elements are the one-dimensional fully marginalized posteriors. The true values of each parameter are marked with X's in the two-dimensional plots and vertical lines in the one-dimensional plots. From left to right, the parameters are $\mc$, $\eta$, $\iota$, distance, right ascension, declination, polarization, and coalescence phase. Note that the geocentric arrival time is omitted because it is marginalized analytically.}
\end{figure*}

All possible sets of one and two-dimensional posteriors over the parameters are shown in Figs. \ref{fig:21091triplot} (\#21091) and \ref{fig:14631triplot} (\#14631). We also display here the posteriors for the intrinsic parameters ($\mc$ and $\eta$). The discreteness of the grid makes determination difficult for the component masses, however, the $\mathcal{M}_c$ and $\eta$ posterior (using the prior in \ref{eqn:mcetaprior}) is consistent with the result shown in Figs. \ref{fig:21091logevid} and \ref{fig:14631logevid}. In both cases, the distinct degeneracy between distance and inclination is clearly shown. In the case of \#21091, a southerly inclination is weakly favored; this being an example of a bimodal distribution where the wrong mode is selected. The better sampled \#14631 does choose the correct mode, but the degeneracy between the two is still quite strong, and the true inclination is nearly exactly face on. Also apparent in both cases is the favoring of moderately off-axis binary configurations which tend to bias the distance measurement towards systematically closer values. The $p(\alpha, \delta)$ posterior, in effect the sky position, resembles the \BS{} skymap which is used by our algorithm to select sample points for these parameters. For \#21091, there are two modes along the triangulation ring, corresponding to a maxima and its mirror image. The sampler was not able to resolve the degeneracy between the points for \#21091\footnote{Though neither did the samplers in \cite{first2years}.}, but the true location of the event lies near one of the maxima on the ring. In the case of \#14631, one mode is suppressed relative to the other, and the true location lies very near the maximum. Also notable is the degeneracy between orbital phase ($\phi$) and polarization angle ($\psi$). The recovered posterior's strong degeneracy  both qualitatively agrees with first principles and suggests further performance improvements (e.g., via direct phase marginalization).

\subsection{Scaling}

For a quasicircular compact binary, it is well-known that the time to coalescence from a given gravitational-wave frequency scales as $t(f) \propto f^{-8/3}$. As the sensitivity of detectors improves at low frequencies, this requires the use of considerably longer waveforms for detection and parameter estimation. For example, the initial LIGO detectors were sensitive down to 40 Hz, while the advanced LIGO detectors could be sensitive down to 10 Hz. To cover the extra low frequency portion would require waveforms that are $\approx 40$ times longer.

Traditional Bayesian parameter estimation is computationally limited by waveform generation and likelihood evaluations. Both of these are linearly proportional to waveform length. Note that the likelihood evaluations involve computing an inner product as in Eq.~\ref{eq:InnerProduct}, which is approximated as a finite sum. The number of points in the sum is determined by the length of the waveform and data being analyzed, which is why the cost of likelihood evaluations scales with waveform length. Therefore, one would expect the cost of Bayesian parameter estimation using a seismic cutoff of $f_{\rm min} = 10$ Hz to be roughly 40 times more expensive than the same analysis using $f_{\rm min} = 40$ Hz.

\begin{figure}
\includegraphics[width=\columnwidth]{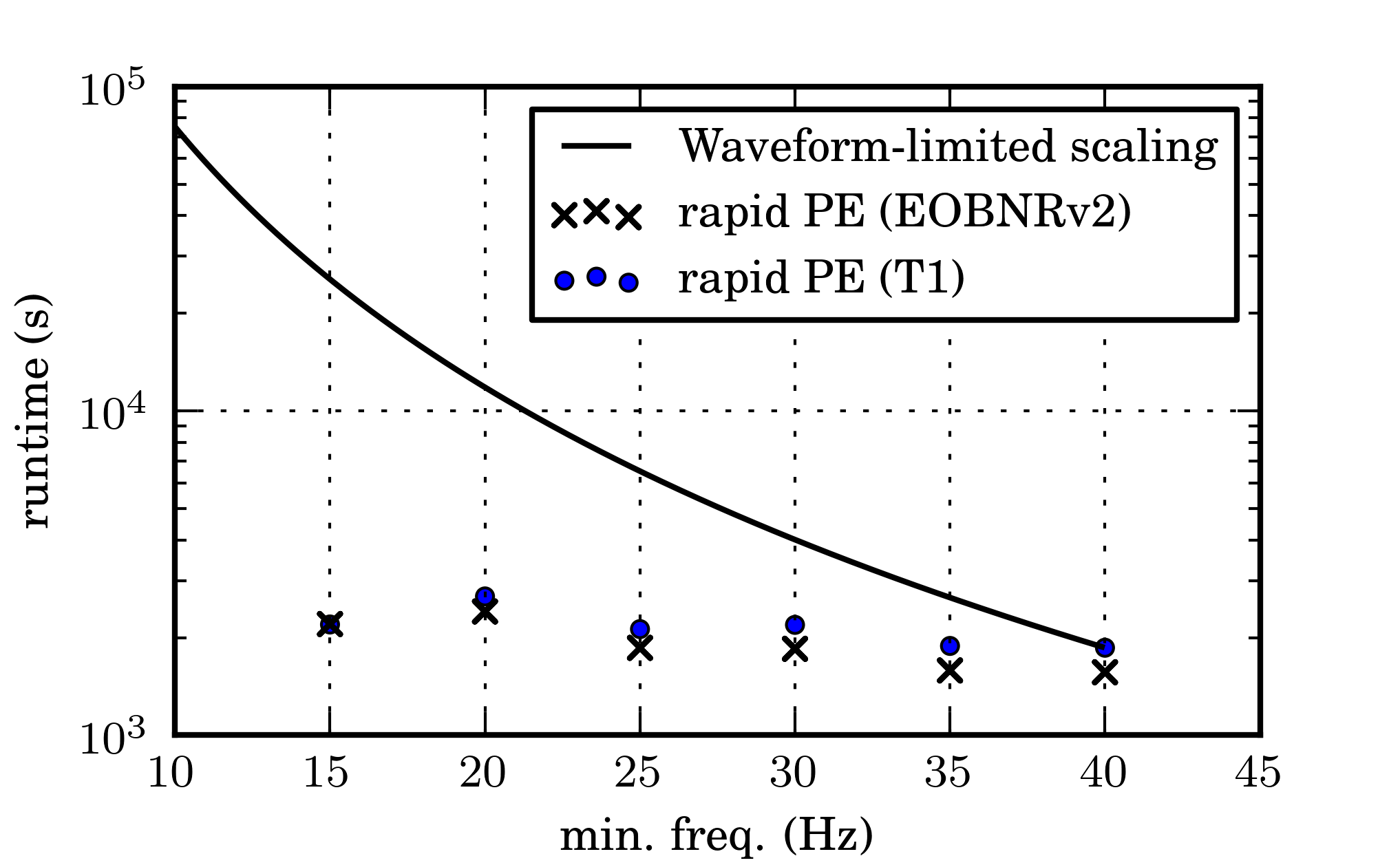}
\caption{\label{fig:fmin_scaling} Points show the runtime of our parameter estimation strategy as a function of the minimum frequency $f_{\rm min}$. For comparison, the solid curve shows the scaling $\propto f_{\rm min}^{-8/3}$ expected if our runtime was proportional to the waveform duration (e.g., runtime proportional to the number of time samples). Waveforms were generated using the standard \texttt{TaylorT1} time-domain code, with $m_1=1.55 M_\odot$ and $m_2=1.23 M_\odot$.}
\end{figure}

The method proposed here is not computationally limited by waveform generation. Recall that for each point in the intrinsic parameter space we compute the waveform and the inner products between the various modes and the data (the assorted $Q_{k,lm}$, $U_{k,lm,l'm'}$, $V_{k,lm,l'm'}$) only once. We then integrate over the extrinsic parameters, which involves evaluating $F_+ + i F_\times$ and the $\Y{-2}_{lm}$'s for different values of the extrinsic parameters. While generating the waveform and computing inner products does scale with waveform duration, this cost is insignificant (even for $f_{\rm min}=10$ Hz) compared to the integration over extrinsic parameters, which is wholly independent of waveform duration. Therefore, as illustrated by  Fig.~\ref{fig:fmin_scaling}. The cost of our method increases only a little as $f_{\rm min}$ is decreased, in contrast to the sharp increase that occurs for waveform-limited techniques such as traditional Bayesian parameter estimation.

\section{Conclusions}
\label{sec:Conclude}

We have demonstrated a viable strategy for low-latency parameter estimation for long-duration binary neutron star signals, using an environment which resembles running conditions during the first advanced LIGO and Virgo observing runs.   

In the era of multimessenger astronomy, rapid and robust inference about candidate compact binary gravitational-wave events will be a critical science product for LIGO, as  colleagues with other instruments  perform followup and coincident observations~\cite{LIGO-2013-WhitePaper-CoordinatedEMObserving}. Low-latency sky localization~\cite{gwastro-skyloc-Sidery2013,LIGO-2013-WhitePaper-CoordinatedEMObserving}, already provides reasonable accuracy by approximate methods like \BS{}, so low-latency parameter estimation will further enable rapid electromagnetic followup and interpretation of candidate gravitational-wave events. 

Motivated by the need for speed, we have introduced an alternative, highly-parallelizable architecture for compact binary parameter estimation. First, by using a mode decomposition  ($h_{lm}$) to represent each physically distinct source and by prefiltering the data against those modes, we can efficiently evaluate the likelihood for generic source positions and orientations, independent of waveform length or generation time.   
Second, by integrating over all observer-dependent (extrinsic) parameters and by using a purely Monte Carlo integration strategy, we can efficiently parallelize our calculation over the intrinsic and extrinsic space.  
Third, to target specific intrinsic (and extrinsic) parameters for further investigation, we ingest information provided by the searches and \BS{}: the trigger masses and estimated sky position.  
Using standard time-domain waveforms in a production environment, we can already fully process one event in less than 1 hour, using roughly $1000$ cores in parallel, producing posteriors and evidence with reproducibly small statistical errors.
By dramatically decreasing the turnaround time for each analysis and by scaling to harness all available resources efficiently, our strategy may significantly increase size and scope of parameter estimation investigations.

Our implementation has bounded runtime -- one hour is the \emph{worst} case -- so we know what resources will be needed to analyze a given NS-NS binary. Moreover, the parallel algorithm can exploit all available computing resources, without need for communication or coordination between jobs, allowing it to operate in computing environments with tightly constrained wallclock time. 

Our algorithm can also be immediately applied to any noise curve and existing time-domain model that provides $h_{lm}$, at any mass, including EOBNRv2HM~\cite{gw-astro-EOBNR-Calibrated-2009} and SEOB~\cite{gw-astro-EOBspin-Tarrachini2012}. 
Finally, by construction our dominant operation count cost is  independent of the waveform's length (or number of basis vectors). Hence, unlike reduced order methods, our code will run in nearly same amount of time now and with full aLIGO-scale instruments with $f_{\rm low}\simeq 10\unit{Hz}$.  

\subsection{Comparison with reduced-order quadrature}

Reduced-order quadrature methods provide an efficient representation of a waveform family and any inner products against it. Other authors have recently proposed prefiltering the data against the reduced-order basis~\cite{gw-astro-ReducedOrderQuadraturePE-TiglioEtAl2014}, achieving significant speedup. For example, using TaylorF2 templates,~\cite{gw-astro-ReducedOrderQuadraturePE-TiglioEtAl2014} claim runtimes of order 1 hour, comparable to our end-to-end time in the high-precision configuration described above.

Our strategy and reduced-order modeling achieve a similar speedup for qualitatively similar reasons: both strategies prefilter the data. In our algorithm, at each mass point, the data is prefiltered against a set of $h_{lm}$, then efficiently reconstruct the likelihood for generic source orientations and distances.  
By integrating the likelihood at each mass point over all extrinsic parameters,  we are dominated by extrinsic-parameter sampling and  hence not limited by waveform generation.

\subsection{Future work}

We first recognize that our sampling strategy, especially the adaptation steps, have some deficiencies, especially in light of earlier literature~\cite{book-mm-NumericalRecipies,peter1978new}. We intend to fortify our current strategy by expanding it beyond the independent product of one-dimensional sampling distributions. High-dimensional sampling distributions can be approximated by kernel density estimators. This will allow us to better capture $n$-dimensional correlations with no appreciable loss of computational speed. This approach will also better allow us to incorporate more detailed information from other stochastic samplers to use as sampling distributions. Improvements to the \BS{} algorithm will also likely allow its direct inclusion as a sampler, instead of just using the data products. Furthermore, the type of parallelization employed here is highly exploitable by GPU accelerated architectures.

While the alternative architecture proposed here is efficient and highly parallelizable over extrinsic parameters, all other parameters are currently suboptimally explored. For example, the algorithm described and implemented here adopts a fixed, low-resolution grid to sample two mass dimensions. All the compact binary specific searches expected to run in the next two years are also capable of providing the specific portion(s) of the template bank that was triggered. We will soon be in a position to develop a more tightly hierarchical approach, and construct our mass grid to resemble those portions.
While the method described here should generalize to a few additional dimensions, it is not yet clear what additional computational resources or architectural changes would be needed to apply our technique to all the intrinsic dimensions (e.g. to include tides and component spins). These higher-dimensional problems are being addressed with Markov Chain or Nested Sampling codes, including tests of GR, models which include non-astrophysical environmental effects~\cite{BayesWave}, and self-consistent electromagnetic and gravitational-wave parameter estimation. In the short-term future, we plan to explore straightforward extension of the intrinsic grid with additional tidal parameters. Also under exploration is using stochastic banks: these types of banks have been used successfully with aligned-spin but otherwise generic binary coalescence searches.
That said, several methods have been proposed for rapid waveform interpolation, including SVD and reduced-order methods. In the long run, we  anticipate being able to perform Monte Carlo integration over intrinsic dimensions as well, without being forced  to adopt the relatively ad-hoc intrinsic/extrinsic split presented here.  

To provide a complete proof-of-principle illustration of our algorithm, we developed an independent production-ready code. That said, the standard \textsc{lalinference} parameter estimation library in general and existing parameter estimation codes (\textsc{lalinference\_mcmc} and \textsc{lalinference\_nest}) could  implement some or all of the low-level and algorithmic changes we describe. For example, MCMC codes could implement our $h_{lm}$-based likelihood, then de-facto marginalize over all extrinsic parameters by strongly favoring jumps at fixed intrinsic parameters ($\lambda$). Any implementation which provides accurate marginalized probabilities (e.g., $\RedLike$) can be parallelized across parameter
space.  
We hope that by combining paradigms, future parameter estimation strategies can reach extremely low latencies, ideally of order a few minutes, when advanced detectors reach design sensitivity. 

\subsection{Acknowledgments}

This work was supported by National Science Foundation awards
PHY~1104371,      
and PHY~1307429.  
We are grateful for computational resources provided by the Leonard E Parker Center for Gravitation, Cosmology and Astrophysics at University of Wisconsin-Milwaukee. We would like to thank Jolien Creighton for useful discussions regarding this paper. We would also like to thank Saeed Mirshekari for a detailed reading of this manuscript. Some of the results in the paper, particularly those derived from \BS{} skymaps make use of the HEALPix\cite{2005ApJ...622..759G} library. This document has been assigned the internal report number LIGO P1500012.

\bibliography{rapid_pe}
\end{document}